\documentclass[11pt,a4paper]{article}
\usepackage{amsmath}
\usepackage{amsfonts}
\usepackage{multirow}
\usepackage{cite}
\usepackage{graphicx}
\usepackage{float}
\usepackage[colorlinks=true,linkcolor=blue]{hyperref}

\newcommand{\be}{\begin{equation}}
\newcommand{\ee}{\end{equation}}
\newcommand{\bea}{\begin{eqnarray}}
\newcommand{\eea}{\end{eqnarray}}
\newcommand{\figref}[1]{Fig.~\ref{#1}}

\newcommand{\R}{\hbox{\upright\rlap{I}\kern 1.7pt R}}

\title{\LARGE{\bf Dynamics of Excited BPS 3-Vortices}}

\author{A. Alonso Izquierdo$^{1,2}$, N. S. Manton$^{3}$,
	J. Mateos Guilarte$^{2}$, M. Rees$^{4}$ and A. Wereszczynski$^{5,6}$
	\\[0.4cm]
	{\normalsize {\it $^{1}$ Departamento de Matematica	Aplicada}, {\it University of Salamanca, SPAIN}}\\
	{\normalsize {\it $^{2}$ IUFFyM}, {\it University of Salamanca, SPAIN}}\\
	{\normalsize {\it $^{3}$ Department of Applied Mathematics and Theoretical Physics}, {\it University of Cambridge, UK}} \\
	{\normalsize {\it $^{4}$ School of Mathematics, Statistics and Actuarial Sciences, University of Kent, UK}}	\\
	{\normalsize {\it $^{5}$ Institute of Theoretical Physics, Jagiellonian University,
			Lojasiewicza 11, Krak\'{o}w, POLAND}}
	\\
	{\normalsize {\it $^{6}$ International Institute for Sustainability with Knotted Chiral Meta Matter (WPI-SKCM2)}}, \\
	{\normalsize {\it Higashi-Hiroshima, Hiroshima 739-8526, JAPAN}}}

\date{}

\begin{document}

\maketitle

\begin{abstract}
We analyze the dynamics of BPS 3-vortex solutions. First, for
unexcited vortices, we study the 2-dimensional moduli space of
centred vortices with $y \to -y$ symmetry, and its metric.
We identify the 1-dimensional subspaces describing the head-on collisions
of equidistant collinear and equilateral triangle configurations,
where geodesic motion results in $90^\circ$ and $60^\circ$ scattering,
respectively. Second, by studying the frequency spectrum of
vibrational shape modes along these subspaces, we explain how the
force-free, geodesic motion is modified by shape mode excitations
into a pattern of chaotic multi-bounce collisions. 
\end{abstract}

%%%%%%%%%%%%%%%%%%%%%%%%%%%%%%
\section{\centering\small MOTIVATION}
%%%%%%%%%%%%%%%%%%%%%%%%%%%%%%

The dynamics of BPS vortices, that is, planar solitons in the
Abelian Higgs model at critical coupling, has
been extensively studied for the last four decades \cite{W, T, SR, Ruback,
	Reb, Samols, JD, Mc, KL, GAR, Hon}, see \cite{MS} for a review. However,
the research has been mainly restricted to the simplest but
still very non-trivial case of scattering of several unexcited
vortices. Unexcited static vortices are BPS solutions, which allows
a semi-analytical treatment of vortex collisions.

An Abelian Higgs vortex configuration has an integer, topological
winding number $n$, so we refer to such a configuration as an
$n$-vortex. BPS solutions satisfy pertinent Bogomolny equations, and
saturate a topological bound on the energy \cite{Bogomolny, Prasad}.
Thus, BPS $n$-vortices are the global energy minima in the winding
number $n$ sector, and they are parametrized by continuous parameters called
{\it moduli}. The manifold of moduli -- the moduli space -- has a
natural curved metric derived from the kinetic energy terms of the field
theory, and unexcited vortex dynamics is accurately
approximated by geodesic motion on the moduli space \cite{NM, Stuart}. This geodesic
approximation has been confirmed by numerical studies of the full
field equations, an example being the $90^\circ$ scattering of two
vortices in a head-on collision \cite{Samols}.

It has only recently been realized that the dynamics of BPS vortices
is far richer and much more interesting than the simplest
force-free geodesic motion. The main change is to pass from unexcited
to excited BPS vortices. Even though the potential energy of a BPS $n$-vortex
solution is independent of its moduli, its discrete spectrum of small
fluctuation frequencies, that is, its spectrum of so-called {\it shape modes}, 
does vary with the moduli. An isolated 1-vortex has a single shape
mode, a radial vibration, but an $n$-vortex solution has several such
modes, and the frequency of each can decrease or increase as the
constituent 1-vortices approach each other \cite{AGMM}. This leads
to a vortex-vortex attractive or repulsive force, which may
completely change the qualitative aspects of the vortex dynamics
\cite{AMMW}. The effect is clearest when the vortices are moving
slowly, and each shape mode amplitude can be treated adiabatically.
The most striking result is the existence of multi-bounce vortex
collisions with chaotic or even fractal-like properties
\cite{KRW}. One possibility is backward scattering, following
an even number of bounces.

The spectral structure can admit an even more radical change. A
shape mode frequency can hit the continuum (mass)
threshold and disappear. This is the spectral wall phenomenon \cite{SW},
clearly observed in a 2-vortex head-on collision
\cite{AMRW}. The existence of the spectral wall
is possible because of a level crossing between two of the
higher-frequency modes when the 1-vortices coincide and the
solution is circularly symmetric. Consequently,
in this case the spectral wall is encountered only after the vortices
coincide. The new phenomena discovered in collisions of two excited
1-vortices can be qualitatively explained using
a simple collective coordinate model, where the 2-vortex
moduli space is enlarged by inclusion of new collective coordinates -
the amplitudes of the shape modes. In particular, from the
spectral flow of the shape mode frequencies, the dynamics acquires
a potential term which generates forces \cite{AMMW}.

In the present paper we continue the analysis of scattering of BPS
vortices, with and without shape mode excitations, in the 3-vortex sector. This is interesting for several reasons.
Firstly, even the force-free, geodesic dynamics of unexcited
3-vortices is not much explored. Very little is known explicitly
about the corresponding moduli space metric. Here, we
analyse the global properties of a 2-dimensional submanifold
of the moduli space, where the centre of mass is fixed at the origin and
the reflection symmetry $y\to - y$ is imposed. This submanifold is still
non-trivially curved. We identify two special 1-dimensional subspaces
within it, and compute the corresponding metric functions along these.

Secondly, the spectrum of shape modes over the moduli space is
more involved for 3-vortices than in the previously studied
2-vortex case \cite{AGMM}. The moduli space is of higher
dimensionality, and there are also more modes. As a result, there
are new scenarios. For example, a spectral wall can occur before or
after the first vortex collision, depending on which mode is excited.
However, all observed dynamical phenomena are again qualitatively
explained by supplementing the geodesic dynamics by a mode-generated potential.

%%%%%%%%%%%%%%%%%%%%%%%%%%%%%%%%%%%%%%%
\section{\centering\small BPS VORTICES}
%%%%%%%%%%%%%%%%%%%%%%%%%%%%%%%%%%%%%%%

The Abelian Higgs model at critical coupling is defined by the
action
\begin{equation}
S[\phi,A]=\int dt \, dx \, dy \Big[ -\frac{1}{4} F_{\mu\nu}F^{\mu \nu} +
\frac{1}{2} \overline{D_\mu \phi}\, D^\mu \phi -\frac{1}{8}
(\overline{\phi}\, \phi-1)^2 \Big] \,,
\label{action1}
\end{equation}
where a complex scalar Higgs field $\phi$ is minimally coupled to
a $U(1)$ gauge potential $A_\mu$.  Here, $D_\mu \phi = (\partial_\mu -i
A_\mu )\phi$ is the covariant derivative, and the electromagnetic
field tensor is $F_{\mu\nu}=\partial_\mu A_\nu - \partial_\nu
A_\mu$. The Minkowski-space metric tensor is
$g_{\mu\nu}={\rm diag}(1,-1,-1),$ with $\mu,\nu=0,1,2$, and we
identify coordinates $(x^0, x^1, x^2)$ with $(t, x, y)$. We will also
use polar coordinates $(x, y) = (r\cos\theta, r\sin\theta)$, and the complex coordinate $z = x+iy$. 

BPS vortices are minima of the static energy functional (in temporal
gauge $A_0=0$)
\begin{equation}
E[\phi,A]= \int_{\mathbb{R}^2} d^2x \Big[ \frac{1}{2} F_{12}^2 +
\frac{1}{2} \overline{D_1\phi}\, D_1\phi + \frac{1}{2} \overline{D_2\phi}\,
D_2\phi+\frac{1}{8} (\overline{\phi}\,\phi-1)^2 \Big] \, ,
\label{energyfunctional}
\end{equation}
obeying the following finite-energy boundary conditions 
\begin{equation}
\overline{\phi}\, \phi|_{\mathbb{S}_\infty^1}=1 \, , \quad 
D_i\phi|_{\mathbb{S}_\infty^1}=0 \quad \mbox{and} \quad
F_{12}|_{\mathbb{S}_\infty^1}=0 \,,
\label{asymptotic}
\end{equation}
where  $\mathbb{S}_\infty^1$ is the circle at spatial infinity. Hence,
the Higgs field gives rise to a map
$\phi_\infty \equiv  \phi|_{\mathbb{S}_\infty^1}: \mathbb{S}^1_\infty
\to U(1)= \mathbb{S}^1$. Each such map can be classified by its topological
winding number, $n \in \mathbb{Z}$, which carries important
physical information. Because $D_i\phi$ vanishes at spatial infinity, 
the vorticity of the
vector potential is the winding number, and via Stokes' theorem,
$\frac{1}{2\pi} \int_{\mathbb{R}^2} d^2 x \, F_{12} = n$, i.e. the magnetic
flux is topologically quantized. The sector with winding
number $n$ can therefore be interpreted as the $n$-vortex sector.

The BPS solutions minimizing $E$ satisfy the first-order
Bogomolny equations \cite{Bogomolny}
\begin{equation}
D_1\phi \pm i D_2 \phi =0 \, , \hspace{0.5cm} F_{12}\pm
\frac{1}{2} (\overline{\phi}\,\phi -1)=0 \, \, , \label{fopdes}
\end{equation}
and saturate the topological bound on
the energy functional, $E[\phi,A]=\pi \vert n\vert$. We
restrict to positive $n$ which requires the upper signs in
(\ref{fopdes}). For given $n$, there exists a
connected, continuous family of BPS solutions characterized 
by $n$ arbitrary, unordered locations in the physical plane, 
$Z_k=x_k+iy_k$, $k=1,\dots,n$, which are the zeroes of the Higgs 
field $\phi$ counted with multiplicity, and simultaneously the 
locations of maximal magnetic field, $F_{12} = \frac{1}{2}$. These
can be understood as locations of $n$ constituent 1-vortices, and they 
provide $n$ complex ($2n$ real) parameters for the solutions, called
moduli. We denote them collectively as $Z$. 
The manifold of moduli can be naturally equipped with a metric
tensor $g_{ij}(Z, {\bar Z}), i,j=1,\dots,n$, forming a curved (K\"ahler)
moduli space \cite{Samols}. The metric functions are found by inserting 
BPS solutions with time-dependent moduli into the kinetic part of
the Lagrangian and performing the space integration (and also imposing
the Gauss Law to ensure that the time-variation of the fields is
orthogonal to a gauge orbit). In the $n=3$ sector,
the moduli space is a complex 3-dimensional manifold with a rather
complicated metric.

The energy degeneracy of all BPS $n$-vortex solutions means
that there are no static forces between the 1-vortex constituents. The
simplest, low-energy dynamics is a force-free motion
along geodesics of the moduli space. Concretely,
the vortex dynamics is governed by the purely kinetic,
collective coordinate Lagrangian 
\be
L[Z_1,\dots,Z_n]=\frac{1}{2} g_{jk} (Z, {\bar Z})
\dot{Z}_j \dot{\bar{Z}}_k, \label{L_eff}
\ee
(where the constant potential energy term $V=\pi|n|$ is omitted).

%%%%%%%%%%%%%%%%%%%%%%%%%%%%%%%%%%%%%%%
\section{\centering\small ZERO MODES -- LINEAR PERTURBATIONS OF A COINCIDENT 3-VORTEX}
%%%%%%%%%%%%%%%%%%%%%%%%%%%%%%%%%%%%%%%

Locally, the existence of $2n$ real moduli is related to the existence
of $2n$ zero modes, i.e. $2n$ independent field variations associated
with variations of the moduli that leave the energy unchanged. For well
separated 1-vortices their existence is obvious. They are
independent translations of the individual 1-vortices. Also fairly
simple is to identify the $2n$ zero modes when the $n$ vortices are
superimposed at the origin, i.e. their locations are coincident and the
configuration has circular symmetry \cite{Burzlaff, AlGarGuil2}.

The circularly symmetric $n$-vortex solution reads
\begin{equation}
\phi^{(n)}(\vec{x}) = f_n(r) \, e^{in\theta} \, ,
\hspace{0.5cm} r A^{(n)}_\theta(\vec{x}) = n\, \beta_n (r) \,,
\hspace{0.5cm} A_r^{(n)}(\vec{x})=0.
\end{equation}
The functions $f_n(r)$ and
$\beta_n(r)$ satisfy the first-order coupled system of ordinary
differential equations (ODE)
\begin{equation}
\frac{df_n}{dr} = \frac{n}{r} f_n(r) [1-\beta_n(r)]
\, , \hspace{0.5cm} \frac{d \beta_n}{dr} = \frac{r}{2n} [1-f_n^2(r)]
\, ,
\label{edo1}
\end{equation}
derived from the BPS equations (\ref{fopdes}), 
together with the asymptotic conditions $f_n(r)\rightarrow 1$ and $\beta_n(r)
\rightarrow 1$ as $r\rightarrow \infty$. Regularity at the origin,
necessary for the energy to be finite, requires $f_n(r) \sim \delta_n r^n$
and $\beta_n(r) \sim \frac{1}{4n} r^2$ for small
$r$, where $\delta_n$ is a constant depending on $n$.

We will denote a zero mode around a BPS vortex solution by
a column four-vector
\begin{equation}
\xi(\vec{x})=\left( \begin{array}{c c c c}a_1(\vec{x}) \,, & a_2(\vec{x})
\,, & \varphi_1(\vec{x}) \,, & \varphi_2(\vec{x}) \end{array} \right)^T,
\end{equation}
where $a_k(\vec{x})$ and $\varphi_j(\vec{x})$ are linear perturbations
of $A_k$ and of the real and imaginary parts of $\phi$.
Exploiting the circular symmetry, in \cite{AlGarGuil2} it was proved
that there exist $2n$ orthogonal zero modes in
$L^2(\mathbb{R}^2)\otimes\mathbb{R}^4$ of the coincident
$n$-vortex of the form
\begin{equation}
\xi_0(\vec{x},n,k) =r^{n-k-1} 
\left( 
\begin{array}{c} h_{nk}(r) \, \sin[(n-k-1)\theta] \\ h_{nk}(r)
\, \cos[(n-k-1)\theta] \\  -\frac{h_{nk}'(r)}{f_n(r)} \,
\cos(k\theta) \\ - \frac{h_{nk}'(r)}{f_n(r)} \,
\sin(k\theta)
\end{array} 
\right), \;\;
\xi_0^\perp(\vec{x},n,k)= r^{n-k-1} 
\left( 
\begin{array}{c} h_{nk}(r) \, \cos[(n-k-1)\theta] \\
-h_{nk}(r) \, \sin[(n-k-1)\theta] \\  -
\frac{h_{nk}'(r)}{f_n(r)} \, \sin(k\theta) \\
\frac{h_{nk}'(r)}{f_n(r)} \, \cos(k\theta)
\end{array} 
\right)  \label{zeromode}
\end{equation}
where $k=0,1,2,\dots,n-1$. 
The radial form factor $h_{nk}(r)$ satisfies the second-order ODE
\begin{equation}
-r \, h_{nk}''(r)+[1+2k-2n\,\beta_n(r)]\,h_{nk}'(r) + r
\,f_n^2(r)\, h_{nk}(r)=0, \label{ode5}
\end{equation}
with the asymptotic conditions $h_{nk}(0)\simeq_{r\to
	0}c_0^{(n,k)}+c_{2k+2}^{(n,k)}r^{2k+2}$ and $\lim_{r\rightarrow
	\infty} h_{nk}(r) =0$, see ref.\cite{AlGarGuil2} for details.

When this circularly symmetric BPS $n$-vortex is perturbed by the
zero mode $\xi_0(\vec{x},n,k)$, the Higgs field near the origin is
\be
\phi(\vec{x},n,k)=  r^k e^{ik\theta} \Big[ \delta_n r^{n-k}
e^{i(n-k)\theta} - \epsilon \,
\frac{(2k+2)c_{2k+2}^{(n,k)}}{\delta_n} \Big] \, \, ,
\ee
so the origin is a zero of $\phi(\vec{x},n,k)$
with multiplicity $k$, and the other $n-k$ zeroes are
vertices of an infinitesimal regular $(n-k)$-polygon, the
$(n-k)$ roots of unity multiplied by an infinitesimal real factor times
$e^{i\frac{\pi}{n-k}}$. Therefore, new zeroes occur at
\be
r(n,k,j)e^{i\theta(n,k,j)}=\Big[\epsilon \frac{(2k+2)\vert
	c_{2k+2}^{(n,k)}\vert}{\delta_n^2} \Big]^\frac{1}{n-k} \,
e^{i\frac{(2j+1)\pi}{n-k} } \hspace{0.5cm}, \hspace{0.5cm}
j=0,1,\dots,n-k-1. \label{zeromodes_zeros}
\ee
The perturbation produced by the zero mode $\xi_0^\perp(\vec{x};n,k)$ 
is similar, but there is an angular shift,
$\theta^\perp(n,k,j)=\theta(n,k,j)-\frac{\pi}{2(n-k)}$. 

\begin{figure}[h]
\centerline{\includegraphics[height=5.5cm]{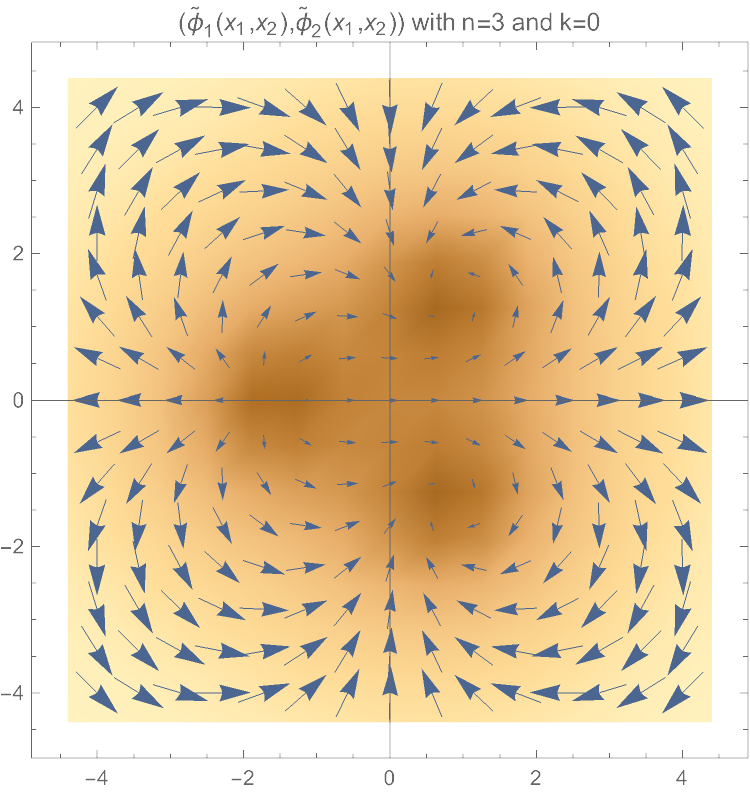}
	\includegraphics[height=5.5cm]{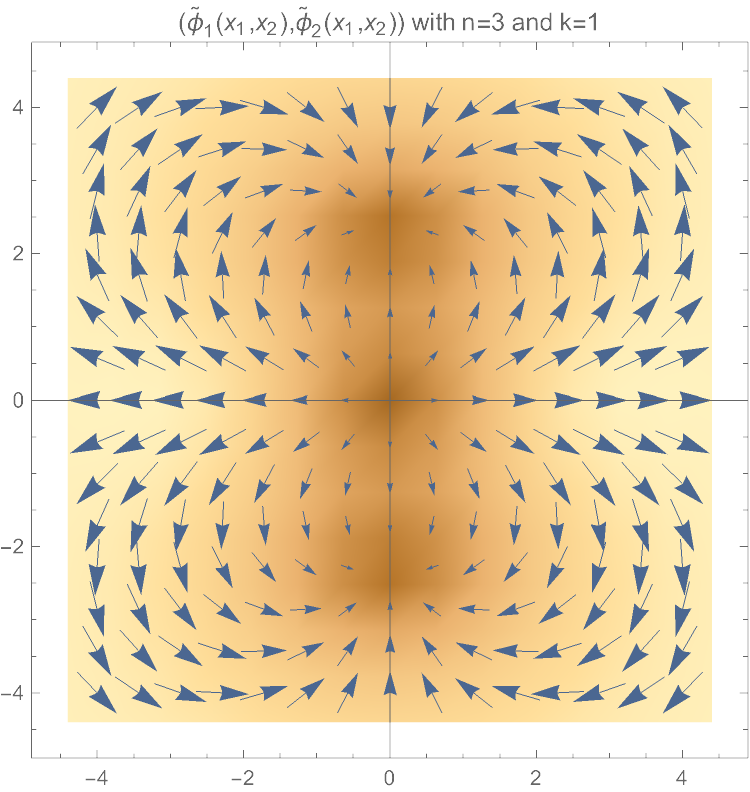}
	\includegraphics[height=5.5cm]{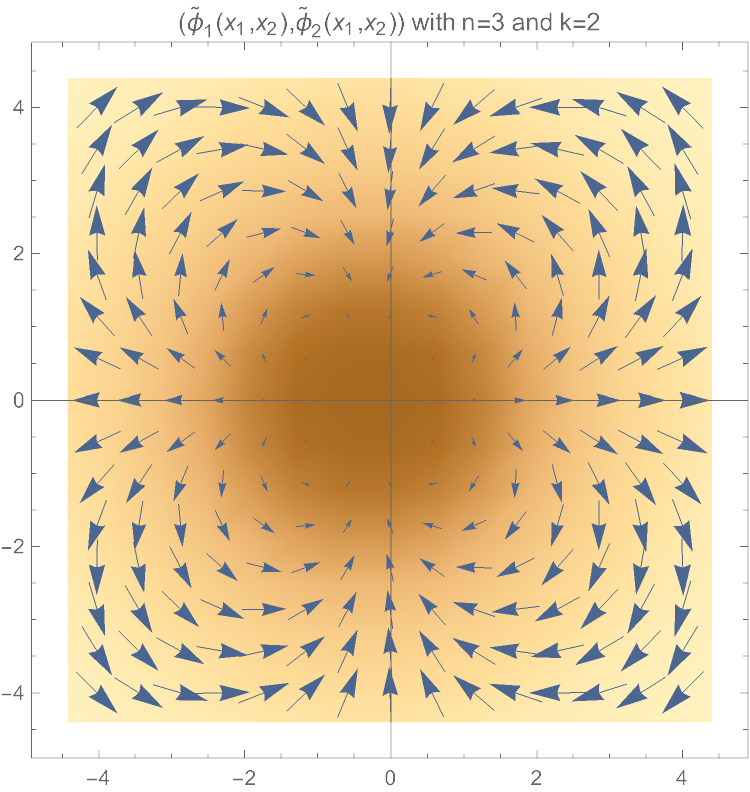}}

\centerline{\includegraphics[height=5.5cm]{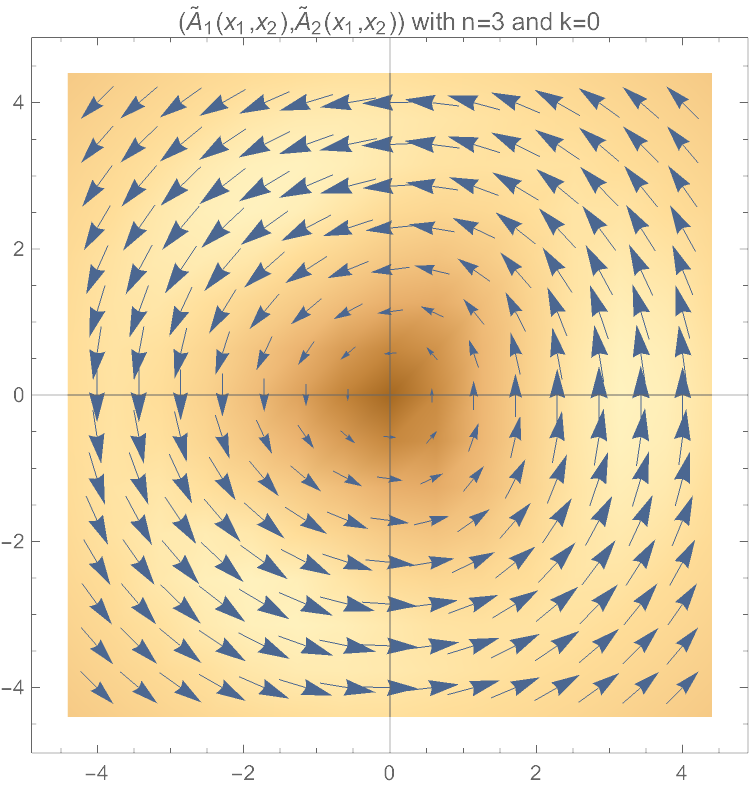}
	\includegraphics[height=5.5cm]{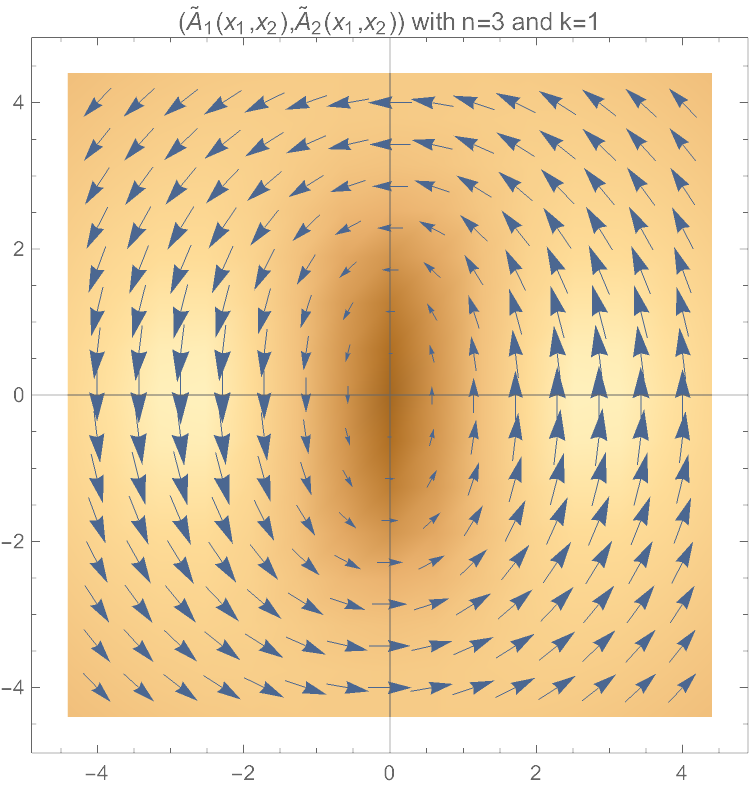}
	\includegraphics[height=5.5cm]{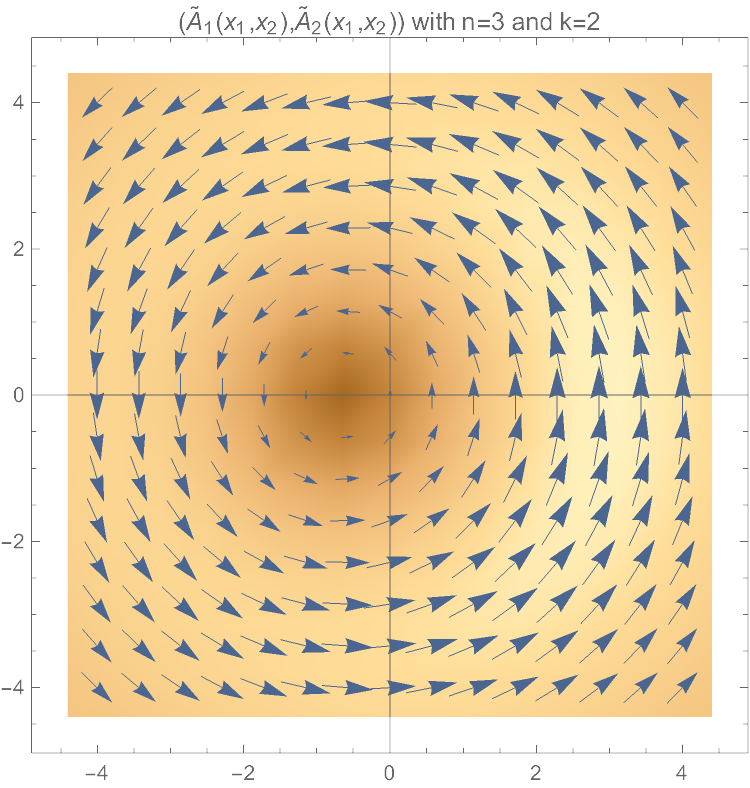}}
	
	\caption{Action of the zero modes $\xi_0(\vec{x},3,k)$ on the circularly symmetric BPS
		3-vortex. Arrows denote orientation of the field (understood as
		two-component vector), while color indicates its absolute value. {\it
			Upper}: the Higgs field. {\it Lower}: the gauge potential.}
	\label{fig:zeromodes}
\end{figure}

In the case of the coincident 3-vortex there are three such pairs of 
zero modes. The $k=0$ mode leads to solutions where 1-vortices are
located at the vertices of an equilateral triangle centred at the origin,
see Fig. \ref{fig:zeromodes} (left panels). The $k=1$ mode leads
to the 1-vortices being collinear and equally spaced, see
Fig. \ref{fig:zeromodes} (middle panels). Finally, the $k=2$ mode splits
the 3-vortex into an $n=1$ and $n=2$ vortex, see Fig. \ref{fig:zeromodes}
(right panels). All  these $\xi_0$ modes enjoy $y\to -y$ symmetry. Furthermore,
the first two modes leave the centre of mass unmoved. This is
not the case for the $k=2$ mode. In summary,
$\xi_0(\vec{x},3,0)$ and $\xi_0(\vec{x},3,1)$ generate all
BPS solutions with $y\to -y$ symmetry and centre of mass fixed
at the origin.
% Strictly speaking, this is true if we add
%configurations generated by $\xi_0^\perp(3,0)$,
%$\xi_0^\perp(3,1)$.

%%%%%%%%%%%%%%%%%%%%%%%%%%%%%%
\section{\centering\small REDUCED 2-DIMENSIONAL MODULI SPACE}
%%%%%%%%%%%%%%%%%%%%%%%%%%%%%%

The moduli space $\mathcal{M}_3$ of the BPS 3-vortex solutions has
local complex coordinates $z_1$, $z_2$ and $z_3$ (which may not be
distinct). As these Higgs field zeroes are unordered, they are best
encoded into the roots of a polynomial \cite{T}
\be
P_3(z) = (z-z_1)(z-z_2)(z-z_3)=z^3+w_2z^2+w_1z+w_0,
\ee
where $z$ is a generic point in the plane and
\be
w_2=-(z_1+z_2+z_3), \;\; w_1=z_1z_2+z_2z_3+z_3z_1, \;\; w_0=-z_1z_2z_3.
\ee
Thus, from the local description, $\mathcal{M}_3=\mathbb{C}^3/S_3$,
where $S_3$ is the three-element permutation group. But the polynomial
description shows that as a manifold,  $\mathcal{M}_3$ is actually
$\mathbb{C}^3$ with the (ordered) complex numbers $w_0,w_1,w_2$
as good coordinates.

Without loss of generality, we set $w_2=0$, fixing the centre of mass
at the origin. This is because the motion of the
centre of mass trivially decouples from the rest of the vortex
dynamics \cite{Samols}. There remains a non-trivial complex 2-dimensional
moduli space $\mathcal{M}_3^{CM}$ parametrized by the polynomials
\be
P^{CM}_3(z) =z^3+w_1z+w_0.
\label{pol0}
\ee
Metrically, this is curved, and still K\"ahler.

We now further truncate $\mathcal{M}_3^{CM}$, by assuming that the
coordinates $w_1$ and $w_0$ take only real values. As a result we
get a real 2-dimensional submanifold $\widetilde{\mathcal{M}}_3^{\, CM}
\subset \mathcal{M}_3^{CM}$. The mapping $y \to -y$ replaces each
root $z_k$ by ${\bar z}_k$, and therefore $w_1$ and $w_0$ by
${\bar w}_1$ and ${\bar w}_0$. $\widetilde{\mathcal{M}}_3^{\, CM}$
therefore parametrizes BPS vortex configurations with the reflection symmetry
$y \to -y$, and it is a complete geodesic submanifold. If the vortex
initial conditions respect this symmetry, then the evolution preserves it.
Because $\widetilde{\mathcal{M}}_3^{\, CM}$ is obtained through a
(non-holomorphic) reflection symmetry, its metric is no longer K\"ahler.

$\widetilde{\mathcal{M}}_3^{\, CM}$ is
generated by the action of two zero modes on the circularly symmetric
solution. To see this, consider the Higgs field configuration
\be
\phi^{(3)}(\vec{x}) + \alpha \xi_0(\vec{x},3,0)+\beta
\xi_0(\vec{x},3,1) \approx \delta_3 r^3e^{3i\theta}
+\alpha\frac{2|c_2^{(3,0)}|}{\delta_3}+\beta\frac{2|c_4^{(3,1)}|}{\delta_3}
re^{i\theta}.
\label{eq:pol01}
\ee
where $\alpha, \beta \in \mathbb{R}$ and $\delta_3= 0.0731242$. If (\ref{eq:pol01}) is rewritten in
terms of $z=r e^{i\theta}$ and the real
coefficients are renamed as
\be
w_1=\beta\frac{2|c_4^{(3,1)}|}{\delta_3^2}, \;\;\;
w_0=\alpha\frac{2|c_2^{(3,0)}|}{\delta_3^2}
\ee
then the $y\to -y$ symmetric polynomial $P_3^{CM}(z)$  is recovered.

It is still useful to consider the locations of the 1-vortices.
To find the roots of the cubic polynomial (\ref{pol0}) we first notice that
its discriminant
\be
\Delta=-4 w_1^3  - 27 w_0^2
\ee
takes real values. There are three possible patterns of roots. 
\begin{enumerate}
	\item If $\Delta>0$ there are three real roots on the $x$-axis, which we denote
	\be
	z_1=2a, \;\;\; z_2=-a+\sqrt{3}\, \tilde{b}, \;\;\; z_3=-a-\sqrt{3} \,
	\tilde{b},
	\label{real}
	\ee 
	where $a, \tilde{b} \in \mathbb{R}$. In this parametrization
	\be
	w_1=-3(a^2+\tilde{b}^2), \;\;\; w_0=-2a(a^2-3\tilde{b}^2),
	\label{realw}
	\ee
	and $w_2=0$ identically. Note that the symmetry
	$\tilde{b} \to - \tilde{b}$ interchanges $z_2$ with $z_3$ and
	leaves $z_1$ untouched.
	
	\item If  $\Delta<0$, there is one real root and two complex conjugate
	roots, which we denote
	\be
	z_1=2a, \;\;\; z_2=-a+i\sqrt{3} \, \tilde{b}, \;\;\; z_3=-a-i\sqrt{3}
	\, \tilde{b}. \label{complex}
	\ee 
	One vortex is still on the $x$-axis while the two others are
	off this axis. Again, $\tilde{b} \to -\tilde{b}$ just exchanges
	vortex labels. This implies $y \to -y$ symmetry. Now
	\be
	w_1=-3(a^2-\tilde{b}^2), \;\;\; w_0=-2a(a^2+3\tilde{b}^2).
	\label{complexw}
	\ee
	
	\item Finally, for $\Delta=0$ we get three real roots where one is a
	double root, denoted
	\be
	z_1=2a, \;\;\; z_2=z_3=-a, 
	\ee 
	and
	\be
	w_1=-3a^2, \;\;\; w_0=-2a^3.
	\ee
	This case continuously connects the two former cases, and if $a=0$
	then all three vortices coincide. 
\end{enumerate}

This verifies that the truncated moduli space
$\widetilde{\mathcal{M}}_3^{\, CM}$ describes 3-vortex solutions
with the reflection symmetry $y \to -y$. All these field
configurations have the property that the generic contours of $|\phi|$, the
Higgs field magnitude, intersect the $x$-axis orthogonally. Consequently, if we
restrict the solutions to the half-plane $y > 0$, then they are BPS
vortices which satisfy Neumann boundary conditions on the boundary $y=0$
\cite{MZ}. The vortex number in the half-plane is 3/2. As discussed in
ref. \cite{MZ}, solutions are classified by the number of half-vortices
on the boundary (a half-vortex is simply a 1-vortex sliced in two by
the reflection line). When considered in the half-plane, the
configurations of type 1 above have three half-vortices on the boundary;
those of type 2 have one half-vortex on the boundary and a 1-vortex
in the interior; and type 3 shows how a pair of half-vortices can
merge on the boundary to form a 1-vortex. 

\begin{figure}
	\centering\includegraphics[width=0.70\columnwidth]{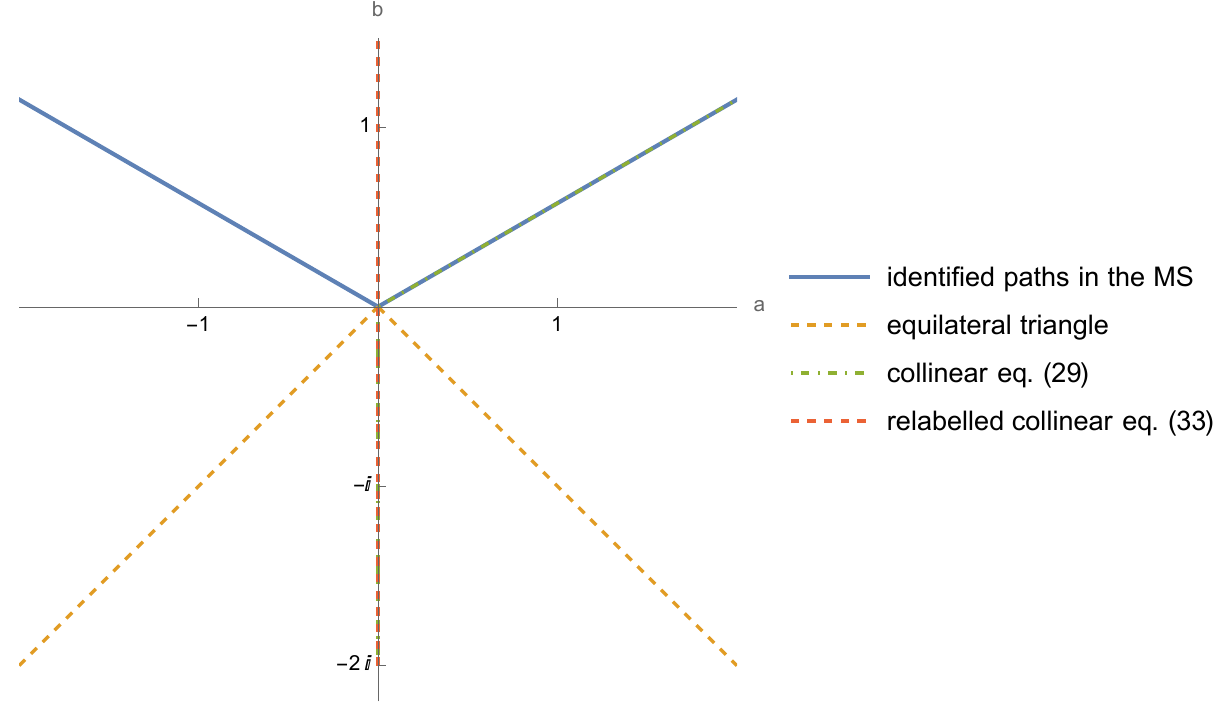}
	\caption{Schematic plot of the 2-dimensional moduli space of $y \to -y$
		symmetric 3-vortex solutions. }\label{fig:MS}
\end{figure}

Because $\tilde{b} \to -\tilde{b}$ only exchanges the
vortex labels, the corresponding BPS solutions are
unchanged. Importantly therefore, in (\ref{real}) and (\ref{complex})
$\tilde{b}$ does not take values in the full real line but rather
in the half-line $\tilde{b} \in \mathbb{R}_+$ (or, of course, equivalently in
$\mathbb{R}_-$). In fact, these cases (together with the boundary case) can be
combined into one expression
\be
z_1=2a, \;\;\; z_2=-a+\sqrt{3}\,b, \;\;\; z_3=-a-\sqrt{3}\,b, \label{unified}
\ee
if we assume that $b \in \mathbb{R}_+ \cup i \mathbb{R}_-\cup
\{0\}$. Here $b \in \mathbb{R}_+$ corresponds to the vortices being located
on the $x$-axis while $b \in i \mathbb{R}_-$ corresponds to two
vortices located off this axis. We schematically plot this moduli space
in Fig. \ref{fig:MS}, which shows that a good coordinate is
$c=b^2 \in \mathbb{R}$, which can be used alternatively in
(\ref{realw}) and (\ref{complexw}).

Also important is the observation that $b=a/\sqrt{3}$ for positive $a$
and $b=-a/\sqrt{3}$ for negative $a$ give the same
solutions. Indeed, they lead to $z_1=2a$ together with
\be
z_2=0, \; z_3=-2a\;\;\; \mbox{and} \;\;\; z_2=-2a, \; z_3=0,
\ee
again just a relabelling, so we have to identify these
lines on the moduli space $\widetilde{\mathcal{M}}_3^{CM}$. Therefore,
while $a \in \mathbb{R}$, the allowed value of $b \in \mathbb{R}_+$
is restricted to $b \leq \frac{|a|}{\sqrt{3}}$. The region
$b > \frac{|a|}{\sqrt{3}}$, i.e. a sector of the $(a,b)$-plane with
opening angle $2 \cdot \frac{\pi}{3}$ needs to be removed from
the moduli space to avoid double counting.

Identifying the half-lines $b = \frac{|a|}{\sqrt{3}}$ (and removing
the sector between them) gives a potentially conical manifold, but to
verify this we need to clarify the asymptotic metric. The
2-dimensional moduli space $\widetilde{\mathcal{M}}_3^{CM}$ is
non-trivially curved in both directions and has no obvious (e.g.
rotational-like) symmetry which further simplifies the metric. However,
the asymptotic metric is flat, and conical, because at large separation,
the kinetic energy is
\be
T = \frac{\pi}{2} \left( |\dot{z}_1|^2 + |\dot{z}_2|^2
+|\dot{z}_3|^2\right) = 3 \pi  \left(\dot{a}^2 + |\dot{b}|^2 \right).
\ee
Thus, we find a flat metric 
\be
ds^2=3\pi\left(d\rho^2 + \rho^2 d\theta\right),
\ee
where 
$\rho^2=a^2+|b|^2$ and $\tan \theta = \frac{|b|}{a}$. Moreover, $\theta
\in \left[ \frac{\pi}{6}, -\frac{\pi}{6}-\pi \right]$. Therefore, we
have indeed a conical deficit angle $2\pi/3$.

%%%%%%%%%%%%%%%%%%%%%%%%%%%%%%
\section{\centering\small 1-DIMENSIONAL GEODESIC SUBMANIFOLDS}
%%%%%%%%%%%%%%%%%%%%%%%%%%%%%%

Further symmetry reductions lead to two 1-dimensional geodesic submanifolds
of $\widetilde{\mathcal{M}}^{CM}_3$. These are derived from the
circularly symmetric, coincident 3-vortex by the separate actions
of the $k=1$ and $k=0$ zero modes, which generate the equidistant collinear and
equilateral triangle configurations respectively. Both subspaces include the coincident 3-vortex.

%%%%%%%%%%%%%%%%%%%%%%%%%%%%%%
\subsection{\small Collinear solutions}
%%%%%%%%%%%%%%%%%%%%%%%%%%%%%%

The first subspace, denoted $\mathcal{N}^{(1)}$, is a truncation in
which the symmetry $y\to -y$ is supplemented by the symmetry $x\to
-x$. These are configurations with
\be
b=\frac{a}{\sqrt{3}},\;\;  a \in \mathbb{R}_+ \;\;\; \mbox{or} \;\;\;
a=0 , \;\; b \in   i \mathbb{R}_-, \label{collinear}
\ee
see the green path in Fig. \ref{fig:MS}. 
In this subspace, one central vortex is always located at the
origin while the two outer vortices are located at equal
distances along the $x$- or $y$-axis, that is, they form
\textit{equidistant collinear} configurations, which we shall refer to
simply as \textit{collinear}. This
subspace is generated by the $\xi_0(\vec{x},3,1)$ zero mode. A
geodesic on $\mathcal{N}^{(1)}$ describes a collinear head-on collision of
three 1-vortices. Initially for $a>0$ they approach along the
$x$-axis, then pass the coincident configuration; finally the
two outer vortices scatter back-to-back along the
$y$-axis, with $a=0$ and $b$ pure imaginary.  Hence, there is
$90^\circ$ scattering (of the outer vortices). Before and after
scattering, 
\be
w_0= 0, \;\; w_1=-12b^2, \;\;\; \mbox{and} \;\;\; w_0= 0, \;\; w_1=-3b^2,
\ee
respectively. $w_1$, or alternatively $c=b^2$, is a good
collective coordinate. 

Note that $w_1$ is a continuous but not a smooth function of $b$ at
$b=0$. This has a simple explanation. Namely, when $b=\frac{a}{\sqrt{3}}$,
\be
z_1=2\sqrt{3}b, \;\; z_2=0, \;\; z_3=-2\sqrt{3}b,
\ee
but when $a=0$,
\be
z_1=0, \;\; z_2=\sqrt{3}b, \;\; z_3=-\sqrt{3}b.
\ee
Hence, the distance between the outer vortices and the origin in
the first case is twice as large as in the second case. This needs to be
corrected by rescaling $b$ by a factor 2 when it is imaginary. $w_1$
evolves smoothly.

We remark that the collinear configurations (\ref{collinear}) are
identical to those with 
\be
a=0, \;\; b \in \mathbb{R}_+ \cup i \mathbb{R}_-\cup \{0\},
\ee
see the red path in Fig. \ref{fig:MS}. 
This corresponds to a smooth $w_1$,
\be
w_0= 0, \;\; w_1=-3b^2.
\ee
However, these configurations are inside the sector of the
2-dimensional moduli space that is excluded to avoid double counting.
Nonetheless, if we restrict ourselves only to  $\mathcal{N}^{(1)}$ this parametrization is valid because it simply corresponds to a relabelling of the vortices.

%%%%%%%%%%%%%%%%%%%%%%%%%%%%%%
\subsection{\small Equilateral triangle solutions}
%%%%%%%%%%%%%%%%%%%%%%%%%%%%%%

The second 1-dimensional subspace, generated by the zero mode
$\xi_0(\vec{x},3,0)$ and denoted $\mathcal{N}^{(2)}$, is a
truncation in which the symmetry $y \to - y$ is supplemented by
cyclic $\mathcal{C}_3$ symmetry, i.e. rotation by
$120^\circ$. The vortices are at the vertices of an
equilateral triangle (simplified occasionally to \textit{triangle})
\be
z_1=2a, \;\; z_2=-a+i\sqrt{3}a, \;\; z_3=-a-i\sqrt{3}a,
\ee
which reverses its orientation (i.e. rotates by $60^\circ$)
as $a$ passes through 0. Hence, $b=ia$. Equivalently, $\tilde{b}=a$
for the configuration (\ref{complex}). Now, 
\be
w_0= -8a^3, \;\;\; w_1=0, \label{w_triangle}
\ee
and although $a$ is an unambiguous coordinate, the good coordinate with smooth
evolution is $w_0$. This geodesic describes a head-on, triangular scattering of three
1-vortices, passing through coincidence and flipping orientation at
$a=0$.

%%%%%%%%%%%%%%%%%%%%%%%%%%%%%%
\subsection{\small Metrics}
%%%%%%%%%%%%%%%%%%%%%%%%%%%%%%

To compute the metric on these 1-dimensional subspaces we apply two
strategies. The first exploits the fact that the subspaces are generated
by appropriate zero modes. As we know the form of
these zero modes in the vicinity of the coincident solution, we can
find the metric in this region. The second approach uses the full
numerical treatment of the slow motion dynamics of the vortices.

In the collinear subspace $\mathcal{N}^{(1)}$, consider a small
perturbation of the coincident 3-vortex along the $\xi_0(\vec{x},3,1)$ zero
mode. Here as a collective coordinate we will use the distance $d$
of the 1-vortices from the origin. Comparing with the previous
parametrization we see that $d=2a$. As already observed, this
coordinate does not provide a good global description of the full
moduli space. However, the missing part can be easily recovered
by continuing $d$ to complex values. On the other hand, the position
of the vortices is a natural observable in the vortex scattering. The
zero mode
\begin{equation}
\xi_0(\vec{x},3,1)= \left(r h_{31}(r) \sin \theta \, , \hspace{0.4cm}
r h_{31}(r) \cos \theta \, , \hspace{0.4cm} -\frac{r
	h'_{31}(r)}{f_3(r)} \cos \theta \, , \hspace{0.4cm} -\frac{r
	h'_{31}(r)}{f_3(r)} \sin \theta \right)^T \, 
\label{zeromodec}
\end{equation}
describes the collinear splitting of the 3-vortex. $h_{31}(r)$ is a
decreasing function so $h_{31}'(r)<0$. The expansion of the Higgs field
of the 3-vortex profile is
\begin{equation}
\widetilde{\psi}^{(3)}(\vec{x}) = f_3(r) e^{3i\theta} - \epsilon \,
\frac{r h_{31}'(r)}{f_3(r)}\,e^{i\theta}  + \cdots \, .
\label{expansion1c}
\end{equation}
A relation between the perturbation parameter $\epsilon$ and the small
distance parameter $d$ can be derived from the zeros of
$\widetilde{\psi}^{(3)}(\vec{x})$. Near the origin,
$f_3(r)\approx \delta_3 r^3$, $h_{31}(r) \approx 1 + c_4^{(3,1)}r^4$
where $\delta_3= 0.0731242$ and the coefficient $c_4^{(3,1)}$ has been numerically estimated
to be $c_4^{(3,1)}\approx -0.036$. Therefore
\begin{equation}
\epsilon = \frac{(\delta_3)^2}{4 \, |c_4^{(3,1)}|} \, d^2 \,,
\label{epsilondrel}
\end{equation}
where the distance of the outer vortices from the origin relates to
the previous parametrization as $d=2\sqrt{3}b$ (with $a=0$).
Putting it all together and assuming an adiabatic dependence
on $d=d(t)$ we have
\be
\Phi(\vec{x},\epsilon,0) = \Phi_0(\vec{x}) + \frac{(\delta_3)^2}{4 \, |c_4^{(3,1)}|} \, (d(t))^2 \,\, \xi_0(\vec{x},3,1)
\ee
where $\Phi=(A_1,A_2,\phi_1,\phi_1)^T$ and
$\Phi_0$ denotes the circularly symmetric solution. Introducing this
expression into the kinetic part of the original Lagrangian, we calculate
\begin{eqnarray}
L &=&  \int d^2 x \, \frac{1}{2} \,  \partial_0
\Phi(\vec{x},\epsilon,0)^T \cdot \partial_0
\Phi(\vec{x},\epsilon,0) \nonumber \\
&=&   \int d^2 x \, \frac{1}{2} \,  \partial_0 \Big[
\Phi_0(\vec{x}) + \frac{(\delta_3)^2}{4 \, |c_4^{(3,1)}|} \,
(d(t))^2 \,\, \xi_0(\vec{x},3,1) \Big] \cdot \partial_0
\Big[ \Phi_0(\vec{x}) + \frac{(\delta_3)^2}{4 \, |c_4^{(3,1)}|}
\, (d(t))^2 \,\, \xi_0(\vec{x},3,1) \Big] \nonumber \\
&=&  \frac{1}{2} \, \,\frac{\,(\delta_3)^4}{4\, |c_4^{(3,1)}|^2} \,
\Big[\int d^2 x \, \xi_0(\vec{x},3,1) \cdot
\xi_0(\vec{x},3,1) \Big] \,  d(t)^2 \partial_0 d(t)
\partial_0 d(t)  \nonumber \\
&=&  \frac{1}{2} \, \, \cdot 0.984019 \, \cdot  d(t)^2 \cdot
\partial_0 d(t) \partial_0 d(t).            
\end{eqnarray}
Thus the metric for small $d$ reads
\be
g_{\rm collinear}(d)= 0.984019 \, d^2 + o(d^2).
\label{gcollin}
\ee

A similar calculation can be made for the subspace $\mathcal{N}^{(2)}$. Now,
the zero mode is
\begin{equation}
\xi_0(\vec{x},3,0)= \left(r^2 h_{30}(r) \sin 2 \theta \, , \hspace{0.4cm}
r^2 h_{30}(r) \cos 2 \theta \, , \hspace{0.4cm} -\frac{r^2
	h'_{30}(r)}{f_3(r)} \, , \hspace{0.4cm} 0 \right)^T \, ,
\label{zeromode5}
\end{equation}
which describes the splitting of the 3-vortex into an equilateral
triangle configuration. As before, $h_{30}'(r)<0$ and the expansion of
the Higgs field is
\begin{equation}
\widetilde{\psi}^{(3)}(\vec{x}) = f_3(r) e^{3i\theta} - \epsilon \,
\frac{r^2 h_{30}'(r)}{f_3(r)}\,  + \cdots \, .
\label{expansion1b}
\end{equation}
Near the origin, $f_3(r) \approx \delta_3 r^3$ and $h_{30}(r) \approx
1 + c_2^{(3,0)}r^2$ with 
$c_2^{(3,0)}\approx -0.258593$. Therefore, the relation between the
infinitesimal parameter $\epsilon$ and the small distance $d$ is now
\begin{equation}
\epsilon = \frac{(\delta_3)^2}{2 \, |c_2^{(3,0)}|} \, d^3 \, .
\label{epsilondrel5}
\end{equation}
Putting it all together with the adiabatic assumption we have
\be
\Phi(\vec{x},\epsilon,0) = \Phi_0(\vec{x}) + \frac{(\delta_3)^2}{2 \,
	|c_2^{(3,0)}|} \, (d(t))^3 \,\, \xi_0(\vec{x},3,0),
\ee
which leads to the reduced Lagrangian 
\begin{eqnarray}
L &=& \int d^2 x \, \frac{1}{2} \,  \partial_0
\Phi(\vec{x},\epsilon,0)^T \cdot \partial_0
\Phi(\vec{x},\epsilon,0) \nonumber \\
&=&  \int d^2 x \, \frac{1}{2} \,  \partial_0 \Big[
\Phi_0(\vec{x}) + \frac{(\delta_3)^2}{2 \, |c_2^{(3,0)}|} \,
(d(t))^3 \,\, \xi_0(\vec{x},3,0) \Big] \cdot \partial_0 \Big[
\Phi_0(\vec{x}) + \frac{(\delta_3)^2}{2 \, |c_2^{(3,0)}|} \,
(d(t))^3 \,\, \xi_0(\vec{x},3,0) \Big] \nonumber \\
&=& \frac{1}{2} \, \,\frac{9\,(\delta_3)^4}{4\, |c_2^{(3,0)}|^2} \,
\Big[\int d^2 x \, \xi_0(\vec{x},3,0) \cdot \xi_0(\vec{x},3,0)
\Big] \,  d(t)^4 \partial_0 d(t) \partial_0 d(t)  \nonumber \\
&=&  \frac{1}{2} \, \, \cdot 0.584678 \, \cdot  d(t)^4 \cdot
\partial_0 d(t) \partial_0 d(t).
\end{eqnarray}
Hence, the metric function for small $d$ is
\be
g_{\rm triangle} (d)= 0.584678 \, d^4 + o(d^4).
\label{gtriangle}
\ee

The behaviour of these metrics close to the coincident solution at
$d=0$ matches the expected scattering angle in head-on collisions. For the
collinear case we have a quadratic zero of the metric. This can be
removed by choosing the correct coordinate $w\sim d^2$, which
corresponds to a $90^\circ$ scattering angle. For the
triangle case the zero of the metric can be removed by $w\sim d^3$
(compare the $a^3$ behaviour in (\ref{w_triangle})). This amounts to
a $60^\circ$ scattering angle.

Equivalently one can extend the domain of $d$. For the collinear case
we assume $d \in \mathbb{R}_+\cup e^{i\pi/2} \mathbb{R}_-$ while for
the equilateral triangle case, $d \in \mathbb{R}_+\cup e^{i\pi/3}
\mathbb{R}_-$. The complex branches realize, respectively, the
$90^\circ$ and $60^\circ$ scatterings of the configurations after
passing through the coincident solution.

The metric around $d=0$ should smoothly (and exponentially fast) approach
its asymptotic, large-$d$ form. Here the metric arises from free motion
of three 1-vortices, each of mass $\pi$, with the imposed symmetry. Therefore,
\be
g_{\rm collinear}^\infty(d) = 2\pi \hspace{0.3cm}\mbox{and} \hspace{0.3cm}
g^\infty_{\rm triangle}(d) = 3\pi
\ee
as $d\to \infty$. 

Our second strategy is to deduce the metric from numerically
determined slow motion of vortices along the 1-dimensional subspaces.
In this case the motion is very well captured by the
collective coordinate model
\be
L[d]=\frac{1}{2} g(d) \dot{d}^2.
\ee
In the numerical simulation, we start with incoming, well separated
vortices each with speed $v_{in}$. Energy is conserved, which allows
us to reconstruct the metric from the actual time evolution of the
distance $d(t)$ of the Higgs field zeroes from the origin, 
\be
g(d)=g^\infty v^2_{in} \frac{1}{\dot{d}^2}.
\ee

\begin{figure}
	\centering
	\includegraphics[width=0.45\textwidth]{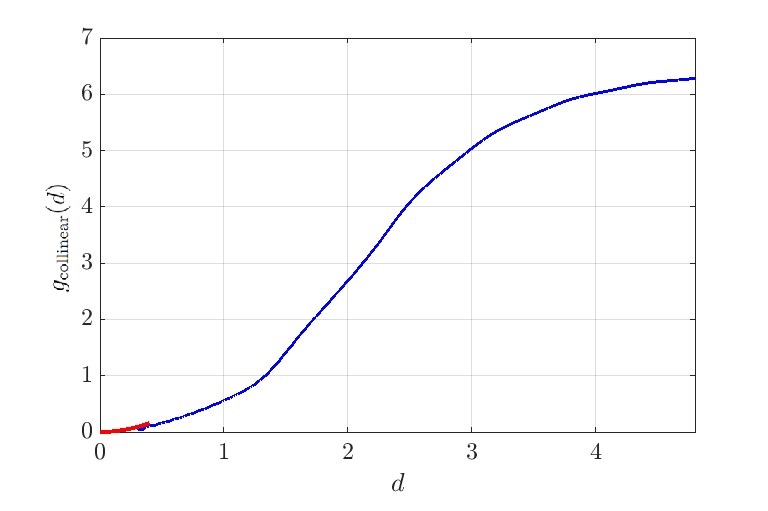}
	\includegraphics[width=0.45\textwidth]{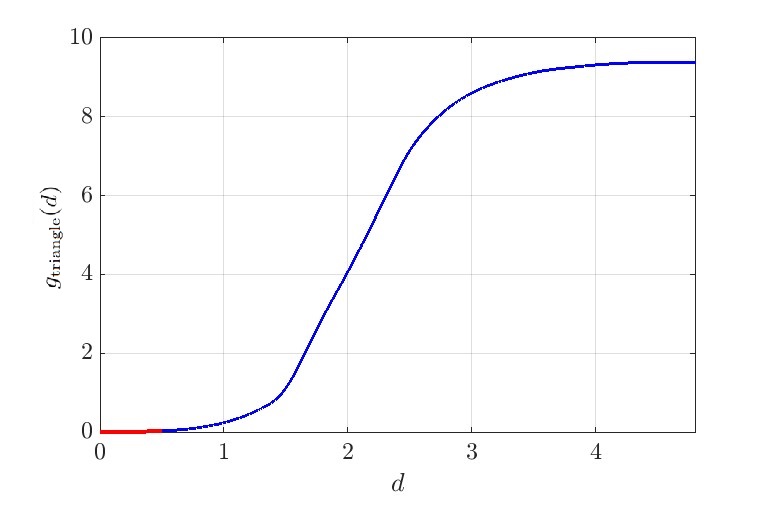}
	\caption{Metric on the collinear subspace $\mathcal{N}^{(1)}$
		(left) and the equilateral triangle subspace $\mathcal{N}^{(2)}$
		(right). The blue curve is the numerically calculated metric
		while the red curve is its analytical approximation near the origin.}
	\label{fig:C0metric}
\end{figure}

In Fig. \ref{fig:C0metric} we plot the resulting metric functions.
Numerically, it is very demanding to compute them when the
vortices are too close together. This is because of problems in
tracking the zeros, as well as because in this
limit $\dot{d}$ diverges. Thus we plot the metrics only for $d>0.3$
(collinear) and $d>0.5$ (triangle), where we
trust the numerics. For smaller $d$ we use the analytical
approximations (\ref{gcollin}) and (\ref{gtriangle}).

%%%%%%%%%%%%%%%%%%%%%%%%%%%%%%
\section{\centering\small SPECTRUM OF VORTEX SHAPE MODES ON
	$\mathcal{N}^{(1)}$ AND $\mathcal{N}^{(2)}$}
%%%%%%%%%%%%%%%%%%%%%%%%%%%%%%

Since the circularly symmetric 1- and 3-vortex solutions are
distinguished points in their moduli spaces, we begin by discussing
the shape modes for these solutions
\cite{Burzlaff, AlGarGuil}. We use notation $\omega_{nk}^2$ for their
(squared) frequencies, where $n$ is the vortex number, and $k =
0,1,2,\dots$ the angular momentum.

\begin{enumerate}	
	\item For $n=1$ there exists only one shape mode, with
	frequency $\omega_{10}^2=0.77747$ and angular momentum $k=0$.
	
	\item For $n=3$ the situation is more complicated. There are three
	shape modes. One a $k=0$ radial mode with
	$\omega_{30}^2=0.402708$ and two degenerate $k=1$ dipole modes with
	$\omega_{31}^2=0.83025$. There are no $k=2$ shape modes, although
	these are theoretically possible, because the effective potential
	minimum is pushed almost up to the continuum. 
\end{enumerate}

The flow of the shape modes for 3-vortices is analysed in a
similar way as for 2-vortices in ref.\cite{AGMM}. Below, we present the
results along the paths generated by the zero
modes $\xi_0(\vec{x},3,1)$ and $\xi_0(\vec{x},3,0)$, that is, along
the 1-dimensional collinear and triangle subspaces $\mathcal{N}^{(1)}$
and $\mathcal{N}^{(2)}$.

%%%%%%%%%%%%%%%%%%%%%%%%%%%%%%%%%%%%%
\subsection{\small Spectral flow in the subspace $\mathcal{N}^{(1)}$ of collinear solutions}
%%%%%%%%%%%%%%%%%%%%%%%%%%%%%%%%%%%%%

\begin{figure}
	\centering
	\includegraphics[height=5cm]{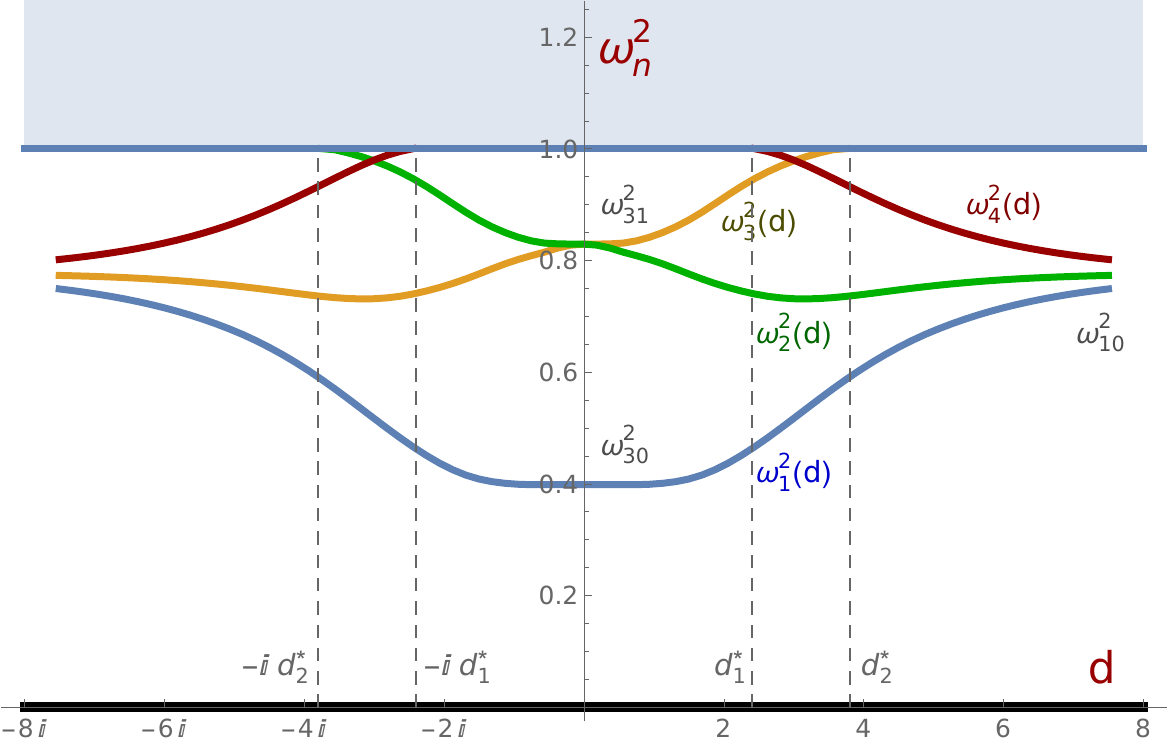} 
	\caption{Spectral flow in the subspace
		$\mathcal{N}^{(1)}$ of collinear vortices. The coordinate
		$|d|$ is the distance from each outer vortex to the vortex
		at the origin.}
	\label{fig:SpectralFlowVortice3b}
\end{figure}

$\mathcal{N}^{(1)}$ is formed by three equidistant collinear BPS
vortices with separation $d$ (and the centre of mass
at the origin). For concreteness, we assume they are located on
the $x$-axis. The spectral flow is depicted in Fig.
\ref{fig:SpectralFlowVortice3b}. The number of shape
modes is the same at $d=0$ and as $d \to +\infty$, but not
all the frequencies interpolate continuously between these limits.

The frequency of the lowest mode $\omega_1^2(d)$ grows from
$\omega_{30}^2$ at $d=0$ to its asymptotic value which is
simply $\omega_{10}^2$. One of the degenerate frequencies at
$d=0$ splits into $\omega_2^2(d)$, interpolating between
$\omega_{31}^2$ and $\omega_{10}^2$. However, the other has
a surprising behaviour. $\omega_3^2(d)$ increases as
$d$ increases and hits the continuum threshold at $d_2^*$, estimated to be
in the segment $d_2^* \in (3.5, 4)$. A further frequency
$\omega_4^2(d)$ descends from the continuum spectrum
at $d^*_1$ in the segment $d^*_1 \in (2, 2.5)$ and then decreases to
$\omega_{10}^2$ as $d \to \infty$. Therefore, in the approximate range
$d \in (2.5,4)$, the spectrum has four discrete frequencies. Moreover,
in this range a level crossing of the
highest two modes takes place (see Fig. \ref{fig:SpectralFlowVortice3b}).

Obviously, the spectrum is symmetrically reflected as we go
to the other part of $\mathcal{N}^{(1)}$ with the vortices
on the $y$-axis, as the outer vortices are simply scattered by
$90^\circ$. The spectrum is combined in one graph in
Fig. \ref{fig:SpectralFlowVortice3b} assuming that
$d \in i \mathbb{R}_-\cup \mathbb{R}_+$. At $d=0$, another interesting
level crossing occurs. The second (third) shape
mode for $d \in \mathbb{R}_+$ continues as the third (second) mode
for $d\in i \mathbb{R}_-$. This level crossing has a crucial
effect on excited 3-vortex collisions, similar to that observed in
excited 2-vortex collisions \cite{AMMW}.

It is important to understand the three almost degenerate shape modes for
1-vortices at large separation. Let $|i\rangle$, $i=1,2,3$ denote
the shape mode of the $i$-th 1-vortex. The lowest 3-vortex mode
(blue curve in Figure \ref{fig:SpectralFlowVortice3b}) is the
in-phase superposition
\begin{equation}
\xi_1= \frac{A}{\sqrt{3}}(|1\rangle+|2\rangle+|3\rangle),
\label{eq:config0}
\end{equation}
whose frequency decreases to $\omega_{30}^2$ as $d \to 0$. The
next (green curve) is the orthogonal superposition
\begin{equation}
\xi_2=  \frac{A}{\sqrt{2}}(|1\rangle-|3\rangle), 
\label{eq:config2}
\end{equation}
whose frequency decreases a little and then increases to
$\omega_{31}^2$ at $d=0$, As already observed, after passing to
imaginary $d$, its frequency increases further until it hits the
mass threshold at $-id^*_2$ and disappears into the continuum. Hence,
this transition to the continuum spectrum happens {\it after} the
constituent 1-vortices coincide. Finally, the third mode (red curve)
is excited by the orthogonal superposition
\begin{equation}
\xi_3= \frac{A}{\sqrt{6}}(|1\rangle-2|2\rangle+|3\rangle),
\label{eq:config1}
\end{equation}
and enters the continuum spectrum at positive
$d^*_1$, i.e. {\it before} the 1-vortices coincide.

%%%%%%%%%%%%%%%%%%%%%%%%%%%%%%%%%%%%%
\subsection{\small Spectral flow in the subspace $\mathcal{N}^{(2)}$ of equilateral triangle solutions }
%%%%%%%%%%%%%%%%%%%%%%%%%%%%%%%%%%%%%

\begin{figure}[h]
	\centering
	\includegraphics[height=5cm]{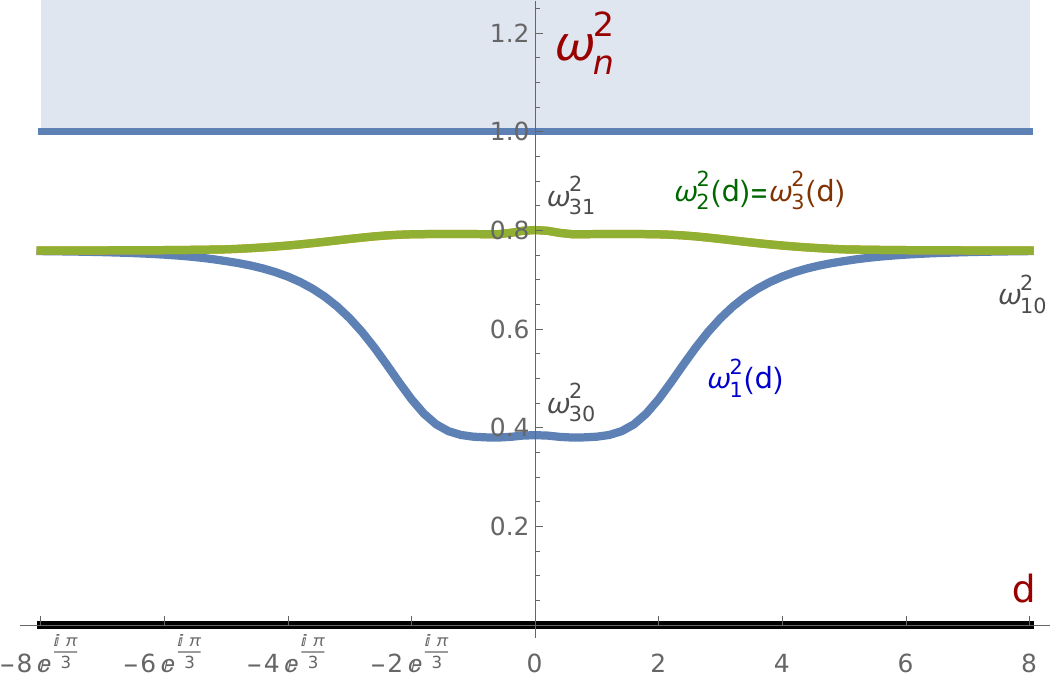} 
	\caption{Spectral flow in the subspace
		$\mathcal{N}^{(2)}$ of equilateral triangle
		configurations. $|d|$ is the distance of the
		vortices from the origin.}
	\label{fig:SpectralFlowVortice3c}
\end{figure}

Here, three BPS 1-vortices are located at the vertices of an equilateral
triangle, each at distance $d$ from the origin. The
spectrum of shape mode frequencies is depicted in Fig.
\ref{fig:SpectralFlowVortice3c}. It interpolates between the spectrum
of the circularly symmetric BPS 3-vortex and the triply-degenerate
frequency $\omega_{10}^2$ of three well separated 1-vortices.
The frequency $\omega_1^2(d)$, starting from $\omega_{30}^2$ at $d=0$,
is almost constant between $d=0$ and $d=2$ and then increases to the
asymptotic value $\omega_{10}^2$. The 2-fold degenerate frequency
$\omega_2^2(d)=\omega_3^2(d)$ is preserved for all $d$, as the relevant shape modes
transform as a doublet of the triangle group $\mathcal{D}_3$. Their frequency
changes little with separation, decreasing from
$\omega_{31}^2=0.83025$ at $d=0$ to $\omega_{10}^2=0.77747$ as $d \to
\infty$. The spectrum is symmetric as $d$ passes through zero, and the vortices scatter by $60^\circ$. 

As before, for well separated 1-vortices, we may excite three
orthogonal modes. The lowest is the in-phase superposition
$\xi_1$ of the three 1-vortex modes, (\ref{eq:config0}),
while the (degenerate) others are the superpositions $\xi_2$ and
$\xi_3$, given by (\ref{eq:config2}) and (\ref{eq:config1}).

%%%%%%%%%%%%%%%%%%%%%%%%%%%%%%%%%%%%%%%
\section{\centering\small UNDERSTANDING THE DYNAMICS OF EXCITED 3-VORTICES}
%%%%%%%%%%%%%%%%%%%%%%%%%%%%%%%%%%%%%%%

The dynamics of $n$ unexcited BPS vortices is captured by the geodesic
flow on the moduli space, with reduced Lagrangian (\ref{L_eff}). This
has as local collective coordinates $Z_j$, $j=1,\dots,n$, the zeroes of
the Higgs field, interpreted as the locations of the constituent
1-vortices on the complex plane. $g_{jk}$ is the moduli
space metric. 

Here we consider the impact of the shape modes on the
dynamics of 3-vortices within the 1-dimensional subspaces
$\mathcal{N}^{(1)}$ and $\mathcal{N}^{(2)}$, parametrizing
head-on vortex collisions with collinear or equilateral triangle
geometry. Each subspace has its own metric, $g_{\rm{collinear}}$ or
$g_{\rm{triangular}}$, giving rise to a simple reduced Lagrangian
\be
L = \frac{1}{2} g(d) \dot{d}^2.
\ee
Around $d=0$, $d$ should be replaced by a new coordinate to avoid
zeroes in the metric function. A better global coordinate is $w=d^2$
for the collinear and $w=d^3$ for the equilateral triangle case,
with $w \in \mathbb{R}$. The coordinate $d$ therefore effectively has a
complex branch. Namely, $d \in \mathbb{R}_+\cup i \mathbb{R}_-$ for
the $\mathcal{N}^{(1)}$ subspace and  $d \in \mathbb{R}_+\cup
e^{i\pi/3} \mathbb{R}_-$ for $\mathcal{N}^{(2)}$.

The force-free geodesic motion can be rather strongly modified if a single
shape mode is excited, because the mode excitation generates an
amplitude-dependent potential, and therefore a force between the constituent
1-vortices. It has been shown that an effective description is
provided by the modified Lagrangian on the moduli space \cite{AMMW}
\be
L=\frac{1}{2} g_{jk} (Z, \bar{Z}) \dot{Z}_j \dot{\bar{Z}}_k + \frac{1}{2}
\dot{C}^2 - \frac{1}{2} \omega^2 (Z,\bar{Z}) C^2,
\ee
where $C$ is the shape mode amplitude and $\omega(Z, \bar{Z})$ its
frequency. Of course, excitation of a higher number of modes, for
example because of level crossing, requires inclusion of more
terms. However, if one restricts to head-on collisions it is
sufficient to keep one mode only and continue its frequency
$\omega$ smoothly through the point of coincidence, see \cite{AMMW}
for a detailed discussion. Thus, to understand the dynamics of
excited 3-vortices in $\mathcal{N}^{(1)}$ or $\mathcal{N}^{(2)}$ we
should analyse the effective model
\be
L_{head-on}=\frac{1}{2} g(d) \dot{d}^2 +  \frac{1}{2} \dot{C}^2 -
\frac{1}{2} \omega^2 (d) C^2,
\ee
where $\omega^2(d)$ is the flow of a single mode frequency
through the whole subspace.

For a given amplitude $C$, there is an attractive/repulsive force
towards the origin if $\omega$ decreases/increases as $d$ decreases.
This is the primary way that the spectral flow of a shape mode affects
the dynamics of excited BPS vortices. The result remains true in
the adiabatic approximation, where $C$ varies, i.e. if the
oscillator behaves as a fast variable, and the vortex motion is
slow, see \cite{AMMW}. 

A further factor determining the dynamics is the
possible existence of a {\it spectral wall} \cite{SW}. This
obstacle arises when a shape mode frequency hits the mass
threshold. If an excited shape mode enters the continuum spectrum at
a point $p_{\rm{sw}}$ on the moduli space, then $p_{\rm{sw}}$ acts as a
source of an additional repulsive force. Depending on the amplitude
$C$ of the mode we find the following scenarios. If the amplitude is
smaller than a critical value, $C<C_{\rm{crit}}$, the repulsion from
$p_{\rm{sw}}$ is overcome by the kinetic energy of the vortices.
They pass through the spectral wall, although they are increasingly
affected as $C$ approaches $C_{\rm{crit}}$. If $C>C_{\rm{crit}}$,
the vortices are back-scattered before reaching the spectral wall,
at a reflection point that becomes more distant from the wall
as $C$ grows. For $C=C_{\rm{crit}}$ the vortices form a long-lived
quasi-stationary state located exactly at $p_{\rm{sw}}$. This spectral wall
phenomenon occurs for a variety of soliton types, e.g. it also has
an impact on the evolution of (near)BPS kinks \cite{SW}, and on
2-vortex dynamics \cite{AMRW}.

Knowing the sign of the force and the position of spectral walls is
sufficient to {\it qualitatively} predict the result of a
scattering. Below we describe their effect on head-on scatterings of
excited BPS 3-vortex solutions within the 1-dimensional subspaces. Our
predictions have been fully confirmed by numerical simulations of the
partial differential equations arising from the field theory.

Our initial configurations were well separated 1-vortices close to
the $d \to \infty$ limit in $\mathcal{N}^{(1)}$ or
$\mathcal{N}^{(2)}$, with one of the available shape modes
excited. The constituents were boosted toward each other along
the subspace, the initial speed of each vortex being always $v_{in}=0.01$.
The evolution was simulated using a $6^{\rm{th}}$-order symplectic
integrator, with time step $\Delta t = 0.01$. A $4^{\rm{th}}$-order
finite difference scheme was implemented for the spatial
derivatives, and we used a lattice of $1001\times1001$ points, with
spatial step $\Delta x = \Delta y = 0.05$. Furthermore, the Lorenz
gauge $\partial_\mu A^{\mu} = 0$ was chosen, and natural boundary
conditions imposed such that the phase of the Higgs field could wind
an integer number of times at the boundary, see \cite{KRW} for details.

The results are presented in the subsections below.
In all plots, $I(0)$ denotes the initial energy intensity
of the mode, defined in terms of the initial mode amplitude
as $I(0)=C^2\omega^2/2$.

%%%%%%%%%%%%%%%%%%%%%%%%%%%%%%%%%%%%%
\subsection{\small Scattering in the subspace $\mathcal{N}^{(1)}$ of collinear configurations}
%%%%%%%%%%%%%%%%%%%%%%%%%%%%%%%%%%%%%

As previously mentioned, the dynamics along $\mathcal{N}^{(1)}$ depends on
which shape mode is excited. We distinguish three possibilities:

\begin{enumerate}
	\item Excitation of the lowest mode \eqref{eq:config0} introduces an
	attractive force between the 1-vortices.  This triggers the resonant
	energy transfer mechanism which results in the appearance of
	multi-bounces with a chaotic, and probably fractal, structure of
	multi-bounce windows. (By a bounce we refer to vortices becoming coincident. A bounce occurs each time 
	$|d|$ hits 0 in Figs. \ref{fig:C0collinear} to \ref{fig:triangle} below.) The explanation is straightforward. During a
	collision, i.e. when the vortices are close to coincident, the
	energy stored in the kinetic motion can be transferred into the
	shape mode. However, it is the balance between the kinetic energy of
	the 1-vortices and the shape mode energy which controls if the
	constituents can separate again. If the amplitude of the mode is too large
	the 1-vortices cannot overcome the attractive force and collide once
	again. At a later time the energy stored in the shape mode is
	transferred back to the kinetic motion and the vortices eventually
	separate. 
	
	\begin{figure}
		\centering
		\includegraphics[width=0.9\textwidth]{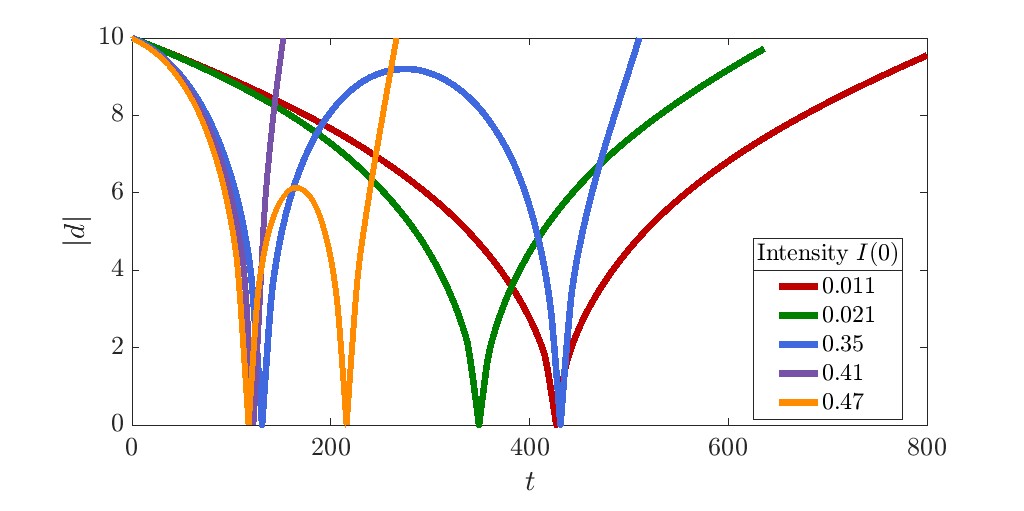}
		\includegraphics[width=0.9\textwidth]{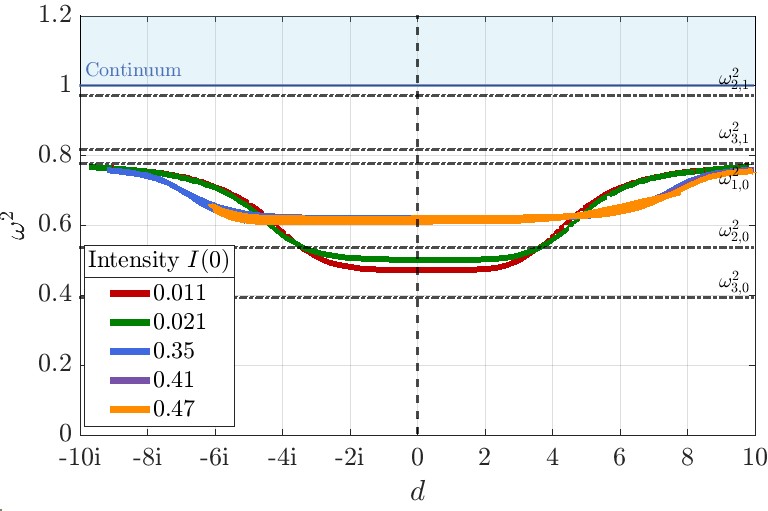}
		\caption{Full field theory numerics for the collinear scattering
			of 3-vortices with the lowest mode \eqref{eq:config0}
			excited: distance $|d|$ of the outer vortices to the
			origin as a function of time (top) and spectral flow as a
			function of $d$ (bottom).}
		\label{fig:C0collinear}
	\end{figure}
	
	In \figref{fig:C0collinear} (top) we plot the distance $|d|$ from the
	outer vortices to the origin. We see a trajectory close to the geodesic
	for small mode intensity $I(0)=0.011$ with only one bounce and the usual
	$90^\circ$ scattering. However, for higher intensities the geodesic
	approximation fails. We see 2-bounce solutions with $180^\circ$
	scattering, where the outer vortices are scattered back to their
	initial  positions. Note that two trajectories with two bounces are
	(chaotically) immersed into 1-bounce trajectories. In
	\figref{fig:C0collinear} (bottom) we also show how the numerically determined
	frequency of the mode varies during the evolution. It very well agrees
	with the spectral analysis. The higher value for the plateau of $\omega^2$
	near $d=0$ observed for higher initial intensities is a numerical artifact
	related to the very rapid motion of the vortices while passing the
	coincident configuration.
	
	\begin{figure}
		\centering
		\includegraphics[width=0.9\textwidth]{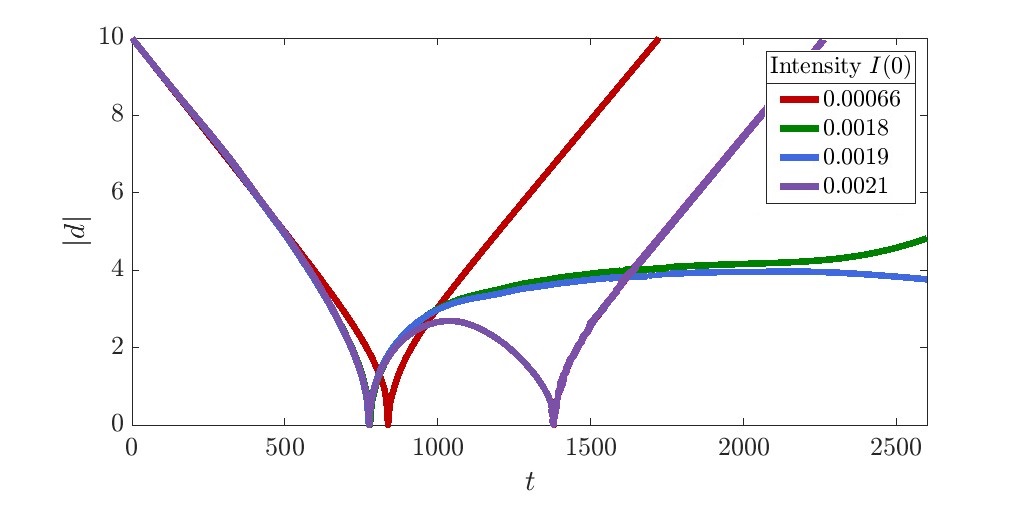}
		\includegraphics[width=0.9\textwidth]{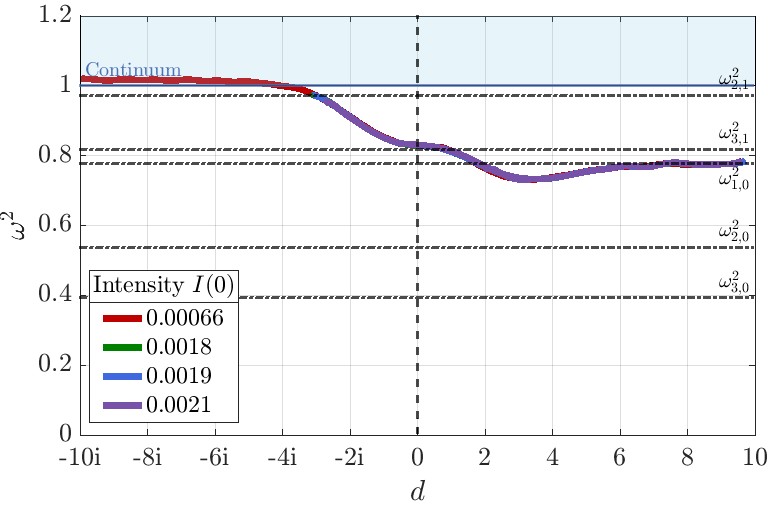}
		\caption{Full field theory numerics for the collinear scattering
			of vortices with the second mode \eqref{eq:config2} excited:
			distance $|d|$ of the outer vortices to the
			origin as a function of time (top) and spectral flow as a
			function of $d$ (bottom).}
		\label{fig:C2collinear}
	\end{figure}
	
	In fact, whenever BPS vortices, or other solitons, have a
	shape mode with the frequency decreasing with the inter-soliton
	distance, we find the analogous resonant energy
	transfer mechanism applying and a chaotic structure of multi-bounce
	windows. This mechanism explains the fractal structure in kink-antikink collisions in the $\phi^4$ model in (1+1)
	dimensions \cite{S, CSW, MORW}. Recently, fractal structure has also been
	identified in the multi-bounces of excited BPS 2-vortices. Depending
	on the number of bounces we have $90^\circ$ (odd) or $180^\circ$ (even)
	scattering.
	
	\item Excitation of the second mode \eqref{eq:config2} introduces a
	repulsive-attractive force. While the vortices are initially well separated
	along the $x$-axis the force is weakly repulsive. If the kinetic energy is
	large enough (or the amplitude of the mode sufficiently small) for
	the vortices to reach coincidence, the force changes
	sign and becomes attractive. This is because of the level
	crossing, with the frequency still growing as the outer vortices
	separate along the $y$-axis. This attractive force can slow the
	vortices and reverse their motion, making them coincide a second time.
	Therefore, depending on the
	amplitude of the excited mode, we can have the following
	scenarios. For large amplitude the vortices may not reach coincidence at
	all. This means no bounces and $180^\circ$ scattering. For a smaller,
	fine-tuned amplitude they can just coincide, but there is still
	$180^\circ$ scattering. For an even smaller
	amplitude they pass through coincidence to the $y$-axis
	but are then attracted back, finally escaping along the
	$x$-axis. This is 2-bounce $180^\circ$ scattering. For
	even smaller amplitudes the outer vortices have enough kinetic energy
	to overcome the attractive force and we observe just one bounce
	and geodesic-like
	$90^\circ$ scattering with the vortices escaping along the $y$-axis.
	
	\begin{figure}
		\centering
		\includegraphics[width=0.9\textwidth]{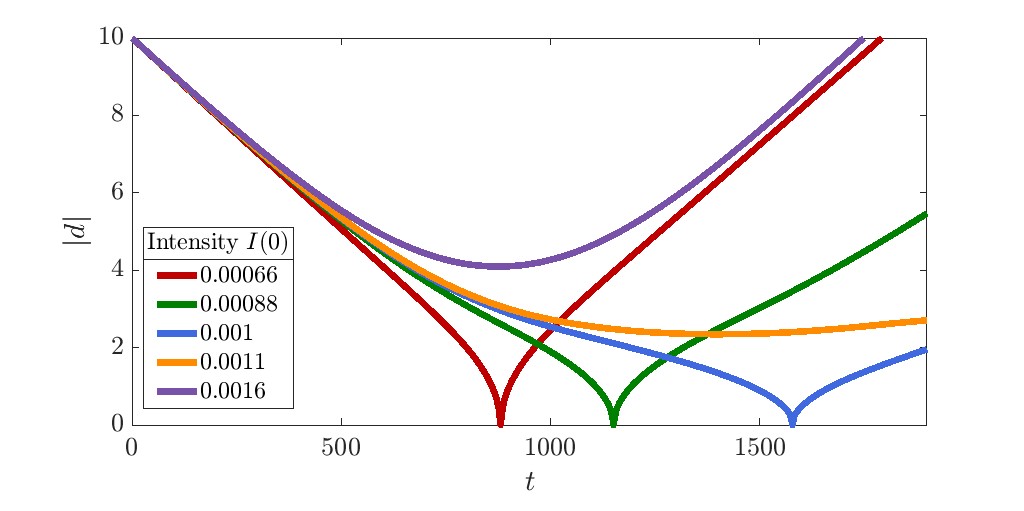}
		\includegraphics[width=0.9\textwidth]{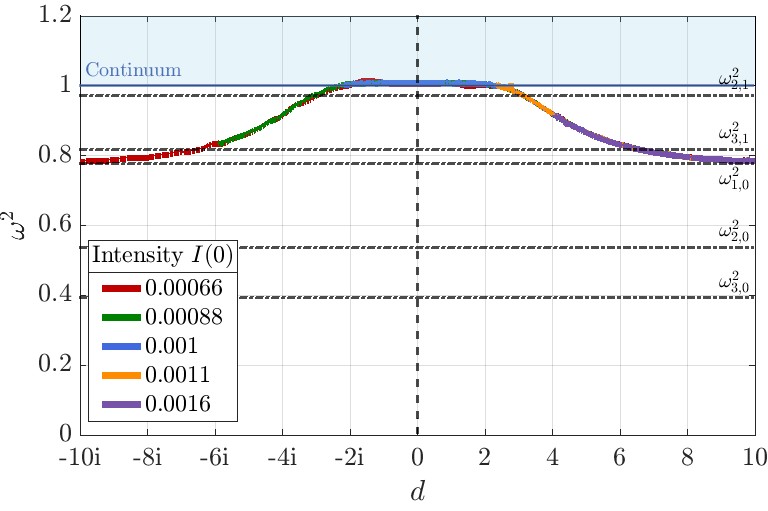}
		\caption{Full field theory numerics for the collinear scattering of vortices
			with the last mode \eqref{eq:config1} excited: distance $|d|$ from
			the outer vortices to the origin as a function of time (top) and
			spectral flow as a function of $d$ (bottom).}
		\label{fig:C1collinear}
	\end{figure}
	
	However, after coincidence, as the mode enters the continuum at
	$id^*$, we expect to see the spectral wall phenomenon.
	This is confirmed in our numerics. In \figref{fig:C2collinear} (top) we
	show the dynamics of $|d(t)|$. Again, for small
	initial intensity $I(0)$ the geodesic motion works very well. However, for
	$I(0)=0.0018$ we find that the vortices form a very long-lived
	quasi-stationary state at separation $d_{\rm{sw}}\approx 4$, close to
	the point on the moduli space where the mode hits
	the continuum, $d^*_2 \in (3.5,4)$. If the intensity is smaller, the
	vortices pass the spectral wall and continue along the $y$-axis. For
	larger intensities, the outer vortices scatter back before reaching the
	spectral wall on the $y$-axis. Finally for
	even larger intensities, they are back-scatterd before reaching
	$d=0$. Hence we see 2, 1 or 0 bounces. In
	\figref{fig:C2collinear} (bottom) we also show the evolving frequency
	of the mode. It is clearly visible that the mode can hit the continuum
	threshold after coincidence. If this
	happens for sufficiently large amplitude the quasi-stationary solution on
	the spectral wall is formed. Note that the case in which the two
	vortices are back-scattered before colliding has not been displayed.
	
	\item Finally, we contrast the dynamics when the last
	shape mode \eqref{eq:config1} is excited. Then, we expect to
	see a spectral wall before coincidence. In this case the force
	is repulsive, and only 0- or 1-bounce collisions can occur. There are no
	2-bounce collisions. Again this is fully confirmed in the numerics,
	see \figref{fig:C1collinear} (top). In this case the stationary
	solution is formed for $I(0)=0.0011$. We clearly see that the location
	of the spectral wall, $d_{\rm{sw}}\approx 2.2$, is within the range
	found in the spectral analysis, $d_1^* \in (2,2.5)$. The
	numerically determined frequency indeed hits the continuum here, see
	\figref{fig:C1collinear} (bottom).
\end{enumerate}

%%%%%%%%%%%%%%%%%%%%%%%%%%%%%%%%%%%%%
\subsection{\small Scattering in the subspace $\mathcal{N}^{(2)}$ of equilateral triangular configurations}
%%%%%%%%%%%%%%%%%%%%%%%%%%%%%%%%%%%%%
%\figref{fig:C2collinear}.
\begin{figure}
	\centering
	\includegraphics[height=5.5cm]{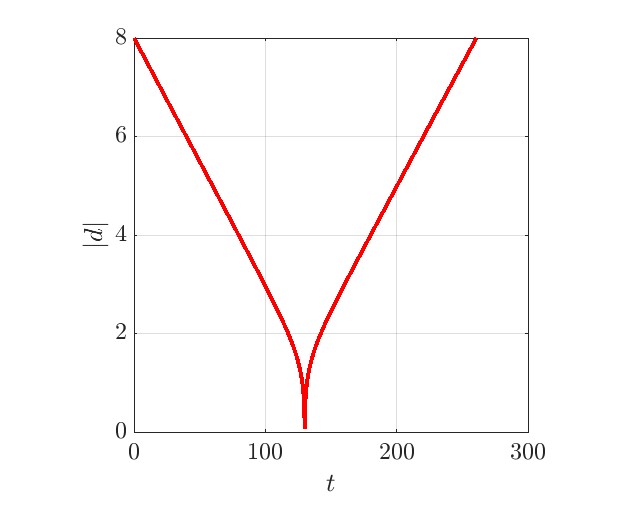}
	\includegraphics[height=5.5cm]{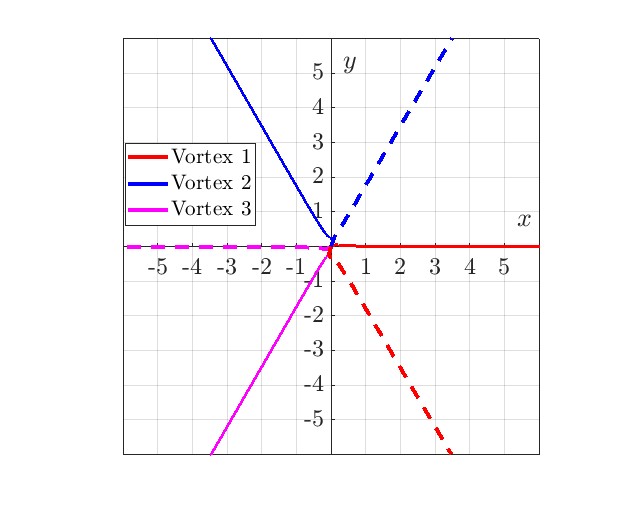}
	\includegraphics[height=8.5cm]{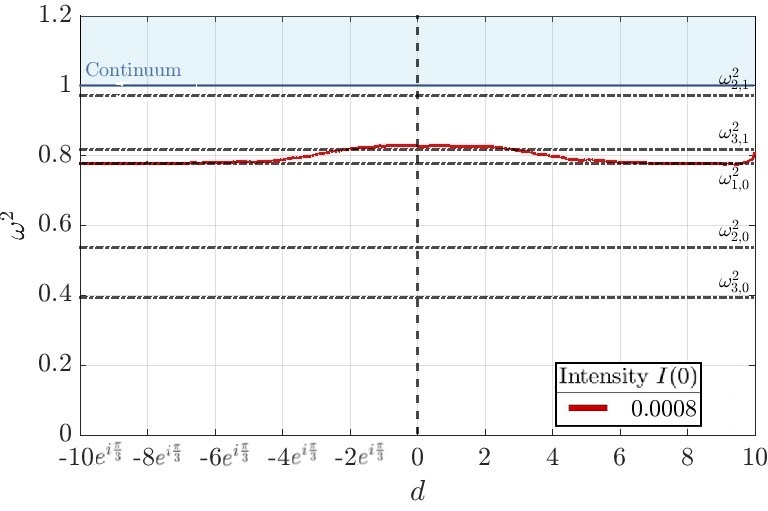}
	\caption{Full field theory numerics for the scattering of vortices
		with the upper mode \eqref{eq:config1} excited in the equilateral
		triangular configuration: Distance $d$ from the vortices to the origin 
		as a function of time (left), geodesic paths of the vortices
		(right), where the dashed lines are the paths after scattering, and
		spectral flow as a function of separation $d$ (bottom).}  
	\label{fig:triangle}
\end{figure}

Excitation of the lowest mode \eqref{eq:config0} in triangular scattering has
a similar effect as in collinear scattering. An attractive force is generated
and the vortices can perform several bounces. Because of the
similarity, we do not plot these trajectories.

Excitation of the two upper modes \eqref{eq:config2} and
\eqref{eq:config1} introduces a very weak repulsive force, which has
an observable effect only if the initial intensity is very large. In
\figref{fig:triangle} we present our numerical results for the
collision with small initial intensity,
$I(0)=0.0008$. In the upper left panel we present the time evolution
of the vortex distance to the origin and trajectories of the
1-vortices in the $xy$-plane during the collision. The geodesic
motion is practically unaffected by the mode
excitation. In \figref{fig:triangle} (bottom) we track the
numerically determined frequency. It exactly follows the pattern
found in the spectral analysis.

%%%%%%%%%%%%%%%%%%%%%%%%%%%%%%%%%%%%%
\section{\centering\small CONCLUSIONS}
%%%%%%%%%%%%%%%%%%%%%%%%%%%%%%%%%%%%%

In the present work we investigated how shape mode excitations affect
the geodesic dynamics of BPS 3-vortices. Due to the
great variety of possible scenarios we restricted attention to head-on
collisions of three 1-vortices with either equidistant collinear
or equilateral triangle geometry.

We also numerically solved the Euler-Lagrange equations with the initial
conditions describing separated, approaching vortices with a shape
mode excited. As the initial state is just a collection of three
1-vortices, there are three orthogonal modes, each being a superposition
of the radial shape mode of the individual vortices. Depending on
which of these superpositions is excited, we saw that the vortices
can attract or repel. Hence, a mode-generated force supplements the
usual geodesic motion.

The lowest mode, asymptotically the in-phase superposition,
always generates an attractive
force. This leads to multi-bounce solutions where vortices scatter
several times. As a result, the $90^\circ$ (collinear) or
$60^\circ$ (triangle) geodesic scattering can be replaced by $180^\circ$
scattering, with an even number of bounces. The number of
bounces depends on the parameters of the collision (initial velocity,
phase and amplitude of the mode) in a rather chaotic way. This matches
results for the BPS 2-vortex \cite{KRW}. For the higher-frequency
modes, the force can be attractive or repulsive, and again the number
of bounces has rather chaotic dependence on parameters.

We also clearly saw the formation of long-lived quasi-stationary
states at certain separations $d_{\rm{sw}}$ in the collinear case. This
is the spectral wall phenomenon, resulting from a mode frequency
hitting the continuum threshold. There are no such states for
the equilateral triangle scatterings.

Looking further, we expect that similar modifications of geodesic
dynamics will occur for other BPS solitons carrying vibrational
shape modes, e.g. BPS vortices
coupled to impurities in the BPS manner \cite{BPS-imp-vortex-1,
	BPS-imp-vortex-2, BPS-imp-vortex-3, BPS-imp-vortex-4,
	BPS-imp-vortex-5, BPS-imp-vortex-6}, and possibly BPS monopoles, see
e.g. \cite{Max-1, Max-2, Max-3}. BPS monopoles do not
host normalized shape modes but instead, infinitely many quasi-normal
modes (Feshbach resonances) \cite{For}. Recently, it has been shown
that Feshbach resonances modify the dynamics of BPS kinks in a very
similar way that the usual shape modes do \cite{AGMC}.
Moreover, in the non-BPS regime (i.e. type I or type II vortices)
a mode-generated force may also alter the usual pattern of attraction or
repulsion, since the modes are very stable against energy
radiation \cite{JJ}.

The fact that the dynamics of vortices can be completely changed by
excitation of shape modes may be important for the dynamics of vortex
strings, considered in various cosmological or astrophysical contexts
\cite{VA, Hind, LISA, St, Hind1, Cop, Hind2}. Indeed, if produced in
phase transitions, such strings will generically be created in excited
states, i.e. with non-zero shape mode amplitudes. Their
interaction, and consequently their annihilation rate and
the gravitational waves they generate will be affected by the excited
modes.

%%%%%%%%%%%%%%%%%%%%%%%%%%%%%%%%%%%%%
\section*{\centering\small Acknowledgments}
%%%%%%%%%%%%%%%%%%%%%%%%%%%%%%%%%%%%%
AAI, JMG and AW acknowledge support from the Spanish Ministerio de
Ciencia e Innovacion (MCIN) with funding from European Union
NextGenerationEU (Grant No. PRTRC17.I1) and Consejeria de Educacion
from JCyL through the QCAYLE project, as well as MCIN Project 1114
No. PID2020-113406GB-I0 and the grant PID2023- 148409NB-I00 MTM. Part
of this research has made use of the high performance computing
systems provided by the University of Kent. MR acknowledges the UK
Engineering and Physical Sciences Research Council (EPSRC) for a PhD
studentship. NSM has been partially supported by STFC
consolidated grant ST/P000681/1.

{\linespread{1}\selectfont
 }

\end{document}